\documentclass[11pt]{article}
\pdfoutput=1 
\usepackage[onehalfspacing]{setspace} % onehalfspacing, doublespacing
\usepackage[ 
	colorlinks = true , 
	urlcolor = blue , 
	linkcolor = black ,
	citecolor = blue 
	]{hyperref}
 
\usepackage[letterpaper,margin=1in]{geometry}
\usepackage{pdflscape}

\usepackage{titlesec}

\usepackage[authoryear,sort&compress,round]{natbib}

\usepackage{amsmath}
\usepackage{amsfonts}
\usepackage{amssymb}
\usepackage{amsthm}

\usepackage{enumerate}
\usepackage{enumitem}

\usepackage{booktabs}
\usepackage[capposition=top,footfont=small,subfloatrowsep=none]{floatrow}
\usepackage[position=bottom]{subfig}
\usepackage{caption} 
\usepackage{afterpage} 

\usepackage{xcolor}
\usepackage{graphicx}
\usepackage{tikz}
\usetikzlibrary{backgrounds,fit,positioning, calc, %
	shapes.geometric, shapes.multipart, shapes, %
	arrows.meta, arrows, %
	decorations.markings, decorations.pathmorphing }
\usepackage[edges]{forest}

\usepackage{multirow}
\usepackage{rotating}
\usepackage{array}
\newcolumntype{L}[1]{>{\raggedright\let\newline\\\arraybackslash\hspace{0pt}}m{#1}}
\newcolumntype{C}[1]{>{\centering}m{#1}}
\usepackage{siunitx}
\usepackage{multirow}
\usepackage{tabularx}
\usepackage{makecell}
\usepackage{enumitem}

\usepackage{appendix}

\usepackage{chngcntr}

\title{Skills and Liquidity Barriers to Youth Employment: \\ Medium-term Evidence from a Cash Benchmarking Experiment in Rwanda}
\date{September 2022}

\author{ 
Craig McIntosh\thanks{\protect\normalsize %
University of California, San Diego, 
\href{mailto:ctmcintosh@ucsd.edu}{ctmcintosh@ucsd.edu}} \,  and
Andrew Zeitlin\thanks{\protect\normalsize Georgetown University, \href{mailto:andrew.zeitlin@georgetown.edu}{andrew.zeitlin@georgetown.edu}}
}

\begin{document}

\maketitle
\thispagestyle{empty}

\vspace{35pt}

\begin{abstract}  

We present results of an experiment benchmarking a workforce training program against cash transfers for underemployed young adults in Rwanda.  3.5 years after treatment, the training program enhances productive time use and asset investment, while the cash transfers drive productive assets, livestock values, savings, and subjective well-being.  Both interventions have powerful effects on entrepreneurship. But while labor, sales, and profits all go up, the implied wage rate in these businesses is low.  Our results suggest that credit is a major barrier to self-employment, but deeper reforms may be required to enable entrepreneurship to provide a transformative pathway out of poverty.

\end{abstract}

\vspace{35pt} 

\begin{singlespace}
\begin{small}
\begin{flushleft}


\textbf{Study Information:} \quad  This study is registered with the AEA Trial Registry as Number AEARCTR-0004388, and is covered by Rwanda National Ethics Committee IRB 114/RNEC/2017, IPA-IRB:14609, and UCSD IRB 161112.  The research was paid for by USAID grant AID-0AA-A-13-00002 (SUB 00009051).  We thank the Education Development Center, GiveDirectly, USAID, and DIL/CEGA for their close collaboration in executing the study; Leodomir Mfura and Melissa Mahoro of Innovations for Poverty Action for outstanding management of the fieldwork; Sarait C\'ardenas-Rodriguez, Grace Han, Atsuhiro Oguri, and Aruj Shukla for excellent research assistance; USAID Rwanda, DIV, and Google.org for funding.  This study is made possible by the support of the American People through the United States Agency for International Development (USAID.) The contents of this study are the sole responsibility of the authors and do not necessarily reflect the views of USAID or the United States Government.  
\end{flushleft}
\end{small}
\end{singlespace}

\clearpage

\cleardoublepage
\pagenumbering{arabic}

\section{Introduction}
%!TEX root=./ms.tex

Sub-Saharan Africa combines a rapidly growing population with low formal-sector employment, meaning that future economic growth will be largely dependent on enhancing productivity in the informal sector \citep{bandiera2022young}.  In this context, few questions have greater long-term import than how best to help the burgeoning young population achieve a successful transition into a productive adulthood \citep{fox2016youth, bongaarts2016development}.  The best means to achieve this are anything but clear, however.  While skills are almost certainly a constraint for a population with the lowest average schooling levels in the world, entrepreneurship and job training programs have an uneven record in contexts with little formal employment \citep{McK21OxREP, kluve2017interventions}.  Credit constraints also certainly play a role, but while a large literature has shown that cash transfers are invested in productive assets in the short term \citep{haushofer2016short, gertler2012investing, blattman2013generating, de2012one}, the ability of transfers to affect durable improvements in productivity is more uncertain \citep{blattman2018long, aizer2016long, hoynes2016long, balboni2019people, baird2019money,  brudevold2017firm}.  More broadly, it is possible that macro-level constraints to demand or to the scope for business expansion fundamentally limit the extent to which the informal sector can provide a pathway out of poverty \citep{la2014informality}.

The effort to help youth make productive transitions into adulthood is inherently a long-term agenda.   We cannot move the needle on this long-term productivity question with interventions that have only palliative, short-term impacts \citep{bouguen2019using}. The existing empirical evidence evaluating labor-market interventions is largely short-term, finding that injections of skills or capital can drive asset ownership, entrepreneurship, and employment over a one to two year time frame.  But few studies have been able to track these outcomes experimentally over a longer period of time (exceptions include \citealp{blattman2020long, blattman2022impacts}).  Serious questions about the durability of the informal sector as a pathway to long-term security have been raised by the Covid-19 epidemic, which has dealt a huge shock to self-employed individuals with income streams vulnerable to lock-downs and without access to employer-based safety nets \citep{mahmud2021household, egger2021falling}.  Since the ability to affect long-term impacts requires the ability to weather shocks \citep{balboni2019people}, the durability of impacts through the Covid-19 epidemic speak both to the dynamics of wealth accumulation and also to the resiliency of different forms of shocks to wealth.

We contribute to this conversation with a study providing a multifaceted window on how best to raise the productivity of vulnerable youth, in this case under-employed 18--25 year olds in Rwanda.  Our study is a randomized controlled trial with one arm providing an intensive year-long vocational training (the \emph{Huguka Dukore Akazi Kanoze} program, henceforth HD), one arm providing unconditional cash transfers (implemented by the U.S. non-profit GiveDirectly, henceforth GD), and an arm that receives both of these interventions at the same time.  Randomization of cash transfer amounts provides the ability to make cost-equivalent comparisons between cash and kind, as well as to form a rich set of counterfactuals for the complementarity arm that receives both interventions.  We follow up with subjects three years after the interventions were completed, and have a permanently untreated control group so the study faces no internal contamination.  Tracking rates in the study were a remarkable 98.6\%, and a relatively even split of male and female subjects allows us to speak to the differential gender dimension interventions.  These study features provide an unusually rich environment in which to consider the medium-term impact of programs that support youth productivity in the African context.

We find evidence that both interventions have sizeable impacts on primary economic outcomes relative to control, with cash transfers out-performing training, and that these impacts persist in the medium term, though they muted relative to the short term and no longer statistically distinguishable from one another.  \citet{mcintosh2022using} reported short-term impacts, finding that HD increased hours worked, monthly income, and productive assets, and that the impacts of cash transfers  were significantly larger at cost-equivalent levels on the latter two domains; moreover, only cash transfers significantly moved consumption outcomes.  The medium-term evidence presented here shows that HD continues to elevate productive hours per week by 3.3, productive assets are almost twice the control group, and an index of business knowledge is higher by 0.25 standard deviations even three year later.  The cash arms led to durable increases in productive assets (between 1.4--3 times the control), subjective well-being, household livestock value, and savings, along with modest and insignificant increases in consumption per capita (10-20\% above the control group).   As was the case in the one-year evaluation results from this study \citep{mcintosh2022using}, we find no evidence of complementarity; the combined arm demonstrates the impacts seen in either arm with no additional benefits arising from them being implemented together.  Most of these outcomes represent a `fade’ of about 50\% relative to the impacts seen in the midline study, meaning that roughly half of the benefit observed after one year is still present more than three years later.  Likely due to this overall diminution in the magnitude of results we find little evidence of significant difference between programs at cost-equivalent level; HD is marginally better in producing business knowledge and other than this we fail to reject differences across arms.

By carefully documenting study participants' time use and entrepreneurship activities, we show that both cash and training drive changes in occupational structure, and---to different extents---drive profitable movements into entrepreneurship that are eroded over the medium term.  Both interventions decrease participation in agricultural wage labor.  The workforce training program weakly pushes individuals into non-agricultural wage labor (5 pp impact), and cash transfers, particularly large ones, drive income in micro-enterprise and particularly non-agricultural self-employment.  These sectoral shifts prove quite constant over time despite the income benefits of the shifts fading after three years.  Both interventions lead to a burst of new business formation over the shorter term; as of the midline the control group had created an average of 0.5 new businesses per person, HD elevated this by 0.2, and the cash arms by 0.5--0.6 new businesses per beneficiary on average.  The rate of new business creation between midline and endline in the control slows to 0.24, and only the GD Large treatment leads to additional new businesses during this interval.  A sizeable fraction of the businesses created at midline die by endline (0.24 in the control group) but this is not more likely in any of the treatment arms.  Many midline businesses are reported as extant but inoperative at endline (0.14 in the control group), and here we see elevated rates for the treatment arms ($\approx 0.15$ greater for the cash arms), suggesting that roughly one third of the businesses created with the cash transfers do not continue to operate three years later.  HD training induced fewer new business to be created, but the marginal firms created under this treatment were more durable. Nonetheless, both programs have sizeable effects on entrepreneurship at endline, with working days, sales, and profits being higher than the control for both training and cash, and profits for the larger cash arms being more than double the control group on average. 

Geography as well as policy responses to the Covid-19 shock provide two critical dimensions of context for our study's results. 
The study takes place largely in rural areas and so many micro-enterprises are typically engaged with agriculture in some way.  Whether such interventions could have more transformative effects in an urban context with larger demand pools remains an open question.  Rwanda is a tightly governed, rapidly growing country.  While in some ways that means that this study likely represents a ``best-case'' scenario for such interventions, it is also the case that the three Covid-19 lock-downs imposed in the two years prior to our endline were unusually strongly enforced, and may have hit small businesses harder than in more loosely governed countries.  The impact of the Covid-19 era on the overall business climate can be seen in our control group: while employment status, consumption, and consumption appear to have been protected over the course of the pandemic, control households have dramatically stripped productive assets, losing approximately 63 percent of the value of the assets they held at midline.
Hence the exigencies of this unusual time are an inextricable part of what this study has to say about long-term impacts.

This study makes several contributions to the literature.  

First, we develop an empirical approach to  \emph{cost-equivalent} comparisons between alternative programs and show its applicability to the study of long-term program outcomes.  Ours is the first study to be able to conduct a rigorously cost-equivalent comparison of two programs over such a long time frame.\footnote{The most common form of benchmarking in the literature is the comparison of food aid to cash aid \citep{leroy2010cash, schwab2013form, hoddinott2014impact, hidrobo2014cash, ahmed2016kinds, CunGioJayXXrestud}.    Efforts to benchmark more complex, multi-dimensional programs to cash include BRAC's Targeting the Ultra-Poor program \citep{chowdhury2016valuing}, microfranchising \citep{brudevold2017firm}, and graduation programs \citep{sedlmayr2017cash}.}  Given the variation in cash transfer amounts we can examine medium-term impacts both allowing the program to change and holding costs constant (cost-equivalence) or allowing program cost to change across modalities (cost-effectiveness).  Finally, because the large cash transfer arm has a cost almost identical to the combined arm that gets both interventions, we can create multiple counterfactuals for the complementarities analysis:  we ask both whether the combination is differentially effective when compared with the sum of its components' treatment effects, and whether the combination is better than the cost of the combination given all in cash.  This suggests several ways of using the ready scalability of cash transfers to create transparent, policy-relevant comparisons.

Second, by providing a clean and well-powered window on the impact of training and cash in a relatively long-term time frame, this study makes a critical contribution to our understanding of the durability of these interventions.
The type of paired classroom and hands-on workforce training that HD Provides is common in workforce programs worldwide, such as the \textit{J\'ovenes en Acci\'on} program in Colombia \citep{attanasio2011subsidizing}, but evidence of impacts of such programs beyond the short term is limited.
Much long-term literature on the impact of cash programs looks at CCTs, which have a pathway to impact either through human capital or the transfers themselves  \citep{fernald200910, barham2014schooling, araujo2017can}.  The long-term literature on income support programs in developed countries illustrates potentially transformative effects on schooling, health, income, and life expectancy  \citep{aizer2016long} and increases in economic self-sufficiency \citep{hoynes2016long}.
Fewer studies have looked at the long-term impact of unconditional transfers in the developing context, but it is far from clear that these impacts are durable, with a number of RCTs showing dissipating long-term benefits \citep{baird2019money, araujo2017can, brudevold2017firm}.  A a long-term study in neighboring Uganda providing cash grants to groups to start businesses showed dissipation of impacts by 9 years from the intervention, with some lasting effects on assets and skilled work \citep{blattman2020long}.  For both short- and long-term studies the training literature has returned mixed results \citep{heckman1999economics, McK21OxREP}, with long-term studies showing some durable impact on formal employment and earnings in the Dominican Republic \citep{ibarraran2019experimental}.  Particularly in the presence of negative economic shocks, it is therefore an open question whether investments in human capital will prove more durable than investments in physical capital enabled by cash transfers.

Finally, the study speaks on a structural level to the constraints that exist to the creation of durable income increases in the informal sector.  On the one hand, our results confirm a literature showing that skills matter \citep{kluve2017interventions}, and that credit constraints matter \citep{beaman2014self}.  They do not suggest there is any special issue at the intersection of credit and human capital constraints that rewards a simultaneous relaxation of these two obstacles.  On the other hand, neither intervention alone, nor the two together, appears capable of delivering a really meaningful escape from poverty over a 3--4 year time frame in this population.  The depressing conclusion of this is that even high-cost interventions may struggle to achieve transformative impacts for vulnerable youth over the longer term, at least in environments buffeted by shocks.  The more expensive interventions in this study cost  approximately \$750 per individual, surely more than most development agencies willing/able to spend, and still do not lead to meaningful decreases in consumption-based poverty after 3.5 years.  A possible reading of this is that we need to think more carefully about interventions that relax constraints on the informal sector as a whole---infrastructure, titling, legal reforms, sector-wide technological investments---rather than investing in individuals while treating these broader capacity constraints as fixed.

The remainder of the paper is organized as follows. Section \ref{s:research_design} presents the design of the experiment and the approach to comparative costing. Section \ref{s:results} presents the main results of the study, including the core experimental results and a comparative cost equivalence and cost effectiveness analysis.  Section \ref{s:aggregation} 4 aggregates short- and medium-term results to provide a summary comparison of cash-flow impacts experienced since treatment.  Section \ref{s:conclusions} concludes.

%--------------------------------------------------------------------------%
\section{Design}\label{s:research_design}
%!TEX root=./ms.tex

\subsection{Interventions}\label{ss:Interventions}

\subsubsection*{Huguka Dukore:  Employment and entrepreneurship readiness training}

\emph{Huguka Dukore}, which means ``get trained and let's work'' in Kinyarwanda, was a five-year project in Rwanda, run by Education Development Center, Inc. (henceforth EDC), and financed by USAID.  Over the lifetime of this project, it provided business-skill training for 40,000 vulnerable youth.  All participants are offered work readiness training; in urban areas the program promotes technical skill and job placement, while in rural areas there is more emphasis on entrepreneurship training and enabling productive self-employment.  Prior to the initiation of this program, the same organization ran a five-year activity in Rwanda, called the Akazi Kanoze Youth Livelihoods Project, that pursued a similar approach.

The implementation of HD studied here is built around three ten-week modules taken sequentially.  The first is called \emph{Work Ready Now!}, providing basic business skills such as accounting as well as emphasizing the  ``soft'' skills hypothesized to be both valuable and transferable across jobs and employment sectors \citep[see][for related evidence]{Cam17science}.\footnote{\emph{Work Ready Now!} consists of eight sub-modules: Personal Development, Interpersonal Communication, Work Habits and Conduct, Leadership, Health and Safety at Work, Worker and Employer Rights and Responsibilities, Financial Fitness, and Exploring Entrepreneurship.} This module consists of 10 five-day weeks of full-day training.  The second 10-week module of HD encourages students to focus on self-employment.  This \emph{Be Your Own Boss} training promotes entrepreneurship and begins to tailor content to the specific sector in which a trainee wishes to focus.  Participants are asked to develop a business idea, identify a concrete market opportunity, outline the business operation and financing, and draw up a business plan.   The third 10-week module is a \emph{Technical Training} course that provides specific skills in an employment area, such as tailoring, hairdressing, carpentry, or beekeeping.

In terms of participation, 86\% of individuals in the HD arm attend the first three weeks of the first module, the definition of compliance on which payment from USAID to EDC is based.  The rates at which individuals complete the different modules are 79\% (Work Ready Now), 69\% (Be Your Own Boss), and 48\% (Technical Training).  After completing these three classroom training components, HD students would typically be placed in an apprenticeship with a local entrepreneur working in the selected employment sector.  39\% of those in the HD arm undertook an apprenticeship during the study period, with the most common placements being in tailoring (53\%) and hairdressing (22\%).

\cite{mcintosh2022using} provide more details of the program, participation rates in the different components of HD, and the specific types of training received.  For our experimental analysis we do not utilize the (endogenous) choices over the specific training received and instead focus on the Intention to Treat (ITT) effect of being offered the bundle that is Huguka Dukore.  

\subsubsection*{GiveDirectly: Household grants program}

The cash arm was implemented by GiveDirectly, a US nonprofit that works extensively in Kenya and has since expanded its implementation capacity to Rwanda, as well as several other Sub-Saharan African countries, Yemen, and the United States.  In Rwanda, they operate using mobile money, first enrolling a household and establishing the phone number through which the transfer can securely be made, then sending the money, and finally following up with the household to verify that transfers were received by the correct person in a timely manner.  If a targeted youth did not themselves have a telephone, they were asked to provide the name and number of a trusted individual to whom the transfer could be sent.  Cash payments for this study were made in two tranches, with the first making up 40\% of the transfer, and the second following one month later and providing the remaining 60\%.  As would be expected, compliance with the cash transfers was essentially perfect.

\subsection{Enrollment and Assignment}\label{ss:enrollment}

The study takes place in three districts; Rwamagana, Muhanga, and Nyamagabe.  Within these districts we selected 13 ``sectors'' (the next geo-political unit in Rwanda below the district), and recruited study participants at the sector level.  To be included in the study youth needed to meet all of the eligibility requirements for \textit{both} implementers, and to have expressed interest in participating in HD by having come to an informational session on the program.  Those youth who met HD's criteria but not GiveDirectly's (meaning that they were not in the poorest two government poverty classifications, Ubudehe 1 or 2) were treated with HD but not included in the study.  

The resulting sample consists of 1,848 individuals.  As shown in Appendix Table \ref{t:Balance}, study participants are just over half female, with an average age of twenty three years, and seven and a half years of education.  Excluding on-farm agricultural labor, exactly a third of the sample reported being employed at baseline.  Median consumption per adult equivalent is 5,879 RWF per month, which in 2018 PPP terms translates to a consumption level of USD 0.66 per day.

Study participants were randomized to treatments in a series of 13 public lotteries, held at the sector level, which were conducted jointly with local officials and representatives of both implementers.
The public lottery mechanism was used given the large sums of money being transferred and the desire to ensure that the entire process was transparent to participants and local officials.  Participants were invited to attend the lottery if they wished to do so but this was not a condition of being included in the study.  If present, participants drew their own treatments via tokens of different colors mixed up in a sack.  We effectively blocked the treatment by ensuring that fixed proportions of tokens for each arm were present in each lottery.  To avoid spillovers within household, for the few households that contained multiple eligible individuals we had them draw a single status for both members, and consequently we cluster standard errors at the household level in the analysis.  

Table \ref{t:treatment_assignment} illustrates the core research design that emerged from the public lotteries.  The four broad arms for the study were a Control, the HD arm, the GD-administered Cash arm, and a Combined arm that received both Cash and HD.  In total, 485 individuals were assigned to HD, 672 to GD, 203 to the Combined arm, and 488 to control.   Within the cash-transfer arm, individuals were randomly assigned to the three bracketing transfer amounts  (\emph{GD-Lower}, receiving \$317.34; \emph{GD-Middle}, receiving \$410.19; and \emph{GD-Upper}, receiving \$503.04), or to the \emph{GD-Large} arm, receiving \$750.  The Combined arm received the Middle GD cash-transfer amount, and the transfers were not conditioned on participation in HD.  In order to minimize tensions over the inequalities generated by the Cash arm, the actual amounts to be transferred were not revealed at the time of the lottery (only the arm to which they had been assigned).  Delivery of the Cash intervention was delayed relative to the start of HD both to allay implementers' concerns that early receipt of cash in the Combined arm might lead participants to drop out of training, and to ensure that individuals in the Combined arm had had time to develop business ideas prior to the receipt of funds.  This meant that HD implementation typically began two months before the first of the cash transfers was made.

\subsection{Cost Measurement}\label{ss:costing}

All costs in this study are calculated from the perspective of USAID, meaning that they include all downstream direct and indirect costs downstream from the funder.  The costing activities consisted of an \emph{ex ante} component in which we attempted to predict implementation costs during the design phase, and an \emph{ex post} costing of what had actually transpired once implementation was complete.  The ex ante costing was used to set the cash transfer amounts, but understanding that this exercise was being conducted under uncertainty, we bracketed the anticipated costs of the HD program with multiple cash amounts so as to allow for adjustment after-the-fact based on the results of the (correct) ex-post exercise.  Our costing followed best practices in this literature and utilized the `ingredients method' \citep{levin2001cost, dhaliwal2012research, levin2017economic, walls2019costing}.  Because HD is a national-scale program serving 40,000 youth, we asked GiveDirectly to synthetically scale up their operations in the costing exercise, providing us with the full operating costs to run a program at that scale for each of the different transfer amounts.  Further details of the costing exercise are provided in \cite{mcintosh2022using}.

 Table \ref{t:costing} shows the evolution of the costing analysis.  The ex-ante exercise arrived at an anticipated HD cost of \$452.47.  Based on the program costs structure underlying this estimate we then bracketed this cost by varying the number of year 1 HD beneficiaries from 8,000 to 12,000, which produced per-capita costs of \$377.05 and \$565.58. The ex-post costing found HD to be less expensive than anticipated, and GD operating costs were slightly higher than anticipated.  This means that the amount USAID spent on HD per \textit{beneficiary} was only \$388.32, while the spending for the GD middle arm was \$493.96.  The inclusion of non-compliance further widens this gap, meaning that USAID cost per \textit{study household} in the HD arm was \$332.27, while in the GD arms it was \$394.93, \$490.99, \$590.41, and \$846.71, respectively.\footnote{Following the terms of USAID's contract with EDC, we attribute no cost to individuals who drop out within the first three weeks of training.}  The combined arm, incorporating compliance with both components of the combined treatment, ended up costing USAID \$840.20 per study individual, an amount similar to the GD Large arm.  These are the numbers used in the cost equivalence analysis.  Because there was no additional implementation in the study sample between midline and endline, for the endline analysis we use the costing numbers from the midline exercise.

\subsection{Surveys and Outcome Measurement}\label{ss:data}

The baselines for the study were conducted during December of 2017 and January of 2018, HD treatment began February 2018, and GD treatment May of 2018.  Study midlines were conducted during July and August of 2019, and this endline activity was conducted in October and November of 2021.   Consequently, the endline is 3 years 8 months after HD started, 3 years 6 months after GD started, and 3 years 10 months after baseline.

The survey exercise tracked the eligible beneficiary as an individual, and administered one survey module to this individual and one to the head of the household in which that person resided (which might also be the beneficiary him/herself).  The study has five primary outcomes.  Employment is a binary measure indicating that the individual spent more than 10 hours in the prior week in paid work or as the primary operator of a micro-enterprise.  Productive hours is the number of hours in the prior week spent in off-farm paid work or in micro-entrepreneurship. Both measures exclude own-farm agricultural work, namely labor put into the farm owned by the household. Monthly income is the total amount earned over the prior month, including enterprise revenue.  Productive asset stocks and household consumption per adult equivalent round out the primary outcomes. 
 
For the midline analysis, we specified three groups of secondary outcomes.  The first group is comprised of components of household wealth, including net non-land wealth, livestock wealth, and the stocks of savings and debt.  
The second included additional measures of beneficiary welfare:  subjective well-being, mental health, and the personal consumption of the beneficiary.  And a third group of outcomes brought together measures of cognitive and non-cognitive skills, measured using Locus of Control, the Big Five index, as well as measures of the aspirations, business knowledge, and business attitudes of the beneficiary.  All monetary outcomes, both primary and secondary, are winsorized at 1\% and 99\% and measured in inverse hyperbolic sine (so that marginal effects can be interpreted as approximating percent changes).

\subsection{Enterprise Data}

Given the strong focus of both interventions on self-employment and the lack of formal employment opportunities in surrounding job markets, the most likely medium for longer-term impacts on income and welfare is enterprises run by the beneficiaries.  To explore this, we examine the results of the two survey modules that were used to measure enterprises.  One of these was based in the household module, and was built to track enterprises primarily run by individuals other than the beneficiary him or herself.  There were relatively few existing household enterprises at baseline, and as we will show, the intervention had quite limited effects on these.  The second module was located in the beneficiary survey, and was built to track businesses either run directly by the beneficiary or to which that individual devoted substantial time or resources.  Because these two instruments were not necessarily administered to the same people or at the same time, it was impossible to design this to completely preclude the possibility of double-counting businesses across these two instruments.  For that reason, we never add together outcomes from these two instruments, instead counting the ``household'' and ``beneficiary'' businesses simply as two different types of entities that are examined separately.

For both household and beneficiary businesses, we collected a number of core enterprise outcomes.  For all extant enterprises we asked whether the business was currently in operation, and if so the number of household and non-household members employed regularly in the business, as well as the number of days that business was operative in a typical month, and the number of customers in a typical month.  We then asked for the typical daily sales on a day when the business is operating, and the total profit the business had earned over the month prior to the survey.  We transform all monetary amounts into US dollars, and then we sum all of these outcomes across all businesses reported at the household level and at the beneficiary level for each beneficiary.\footnote{As in the rest of the analysis, non-binary outcomes are Winsorized at 99\% and monetary outcomes are inflation adjusted to make them real midline US dollars.}  These totals are then merged back into the experimental dataset (of all beneficiaries, whether they run a businesses or not), missing values are replaced with zeros, and we run standard analysis of the effects of the interventions as elsewhere in the study.  These impacts can then be interpreted as average effects of the interventions on the total of each variable in a manner that combines extensive margin impacts on the existence of a business with intensive margin impacts on the size of existing businesses.

We also attempted to panel track these entities, thereby generating the ability to answer questions about the birth and death of specific enterprises over time.  Because the household enterprises are less dynamic the interest in this exercise applies mostly to the beneficiary businesses.  Using panel tracking, we can define a number of different variables at the enterprise level that pertain to the extensive margin.  For both midline and endline we can define `new firms' that had not existed in previous waves and are born in that round.  Then, for the midline, we classify the existing firms that we observe in that round into three categories; ``will survive'', ``will be inoperative'' (in endline the respondent says the firm still exists but is not currently operative), and ``will die'' (firm no longer exists at endline).  As above, because the sample of individuals with firms is strongly endogenous to the treatments, it is unattractive to analyze these outcomes in rates at the firm level; instead we total them at the beneficiary level, replacing missing values with zeros for the new businesses, and so create outcomes that can be analyzed in a standard experimental context that are ``number of new firms''.  Among the endogenous sample of firms at midline, we can then examine what happens to those firms by endline as a function of treatment status.

\subsection{Attrition and Balance}

We attempted to follow up with all study beneficiaries at endline, 46 months after baseline, regardless of whether they had been successfully tracked at midline or not.  We followed the beneficiary youth as an individual, and considered the ``household'' to be the place in which that individual was resident at the time of endline even when that differed from the baseline household.  The survey teams initiated a first phase of tracking where they attempted to find all individuals who had moved within their home districts or had gone to Kigali, the capital.  We had originally intended to randomly sample from the remaining individuals not found by this process to conduct an ``intensive tracking'' exercise, as we did in the midline, but the original tracking was so successful, and the remaining sample sufficiently small, that in the end we simply intensively tracked everyone in the study.  This intensive tracking phase involved sending an enumerator to speak with them in person if they were located anywhere in Rwanda or Uganda (where IPA has a sister office and therefore could easily mount in-person surveys), and then conducting a phone survey with anyone who could not be located through the above means or who had migrated to a different country. 

For a potentially highly mobile sample of youth, our tracking was remarkably successful: in the end we managed to survey 98.6\% of all baseline youth.  This is higher than we had anticipated and may be due to the advent of the Covid-19 shock in the interim that made migration and work away from home more difficult, thereby keeping the study sample less mobile than they might have been in business-as-usual circumstances.  Of the 1,848 baseline individuals, at endline we found 8 in jail, 8 passed away, 3 mentally ill, 1 in military training, 4 refused the endline survey, and 2 individuals that we failed to find, for a total of 26 baseline individuals who were not included in the endline survey.  

Table \ref{t:Attrition_simple} analyzes differential attrition by arm.  Overall, as is unsurprising with such a high tracking rate, we do not find differential attrition by arm.  The one exception is in the GD Large arm, where we succesfully tracked all 178 individuals assigned to this arm, and the resulting tracking rate of 100\% is significantly different from the control rate of 98.6\%, although clearly the absolute difference is very small in magnitude.  Table \ref{t:Attrition} looks for signs of differential determinants of attrition by regressing baseline covariates on a dummy for whether the individual was successfully tracked at endline.  Only for one covariate do we see any signs of differences, namely that we were least successful in tracking those individuals who were wealthiest at baseline in terms of consumption.  Nonetheless, once adjusted for multiple inference across outcomes we find no evidence of overall tracking differentials, meaning that the endline sample is representative of the baseline universe.

We can then examine the balance of the experiment using the attrited endline sample that will be used for analysis.  Table \ref{t:Balance} shows an exceedingly well-balanced sample, with not a single covariate significant for any arm once adjusted for multiple inference.  The endline sample therefore appears to provide clean internal validity and a remarkably well-tracked sample given the duration of the effects we estimate here.

%--------------------------------------------------------------------------%
\section{Results}\label{s:results}
%!TEX root=./ms.tex

We present the results of the study in six steps.  
First, we present ITT effects on primary and secondary outcomes. 
Second, we show that both intervention types---but particularly the cash-transfer treatments---led to the creation of businesses in the short term, and that entrepreneurship activity remains substantially elevated levels of beneficiary operation even 3.5 years after treatment.    
Third, we test formally for complementarities, finding no evidence of technological complementarities between these interventions either at midline or endline.  
Fourth, we use the random assignment to alternative cash transfer amounts and a linear dose-response function to make a cost-equivalent comparison between the two intervention modalities.    
Fifth, we contrast this approach to a traditional cost-effectiveness analysis, comparing arms in impacts-per-dollar terms, and we interrogate assumptions of heterogeneity and spillovers that are potentially important to a projected scale-up of these interventions.  
And sixth, we examine the incidence of Covid-19-related shocks and their potential to explain the attenuation of impacts.

\subsection{ITT Results on Primary and Secondary Outcomes}

To begin, we estimate ITT impacts of assignment to HD or cash transfers on primary and secondary outcomes.  For comparison, we reprised midline results reported in \citet{mcintosh2022using}, presenting these alongside new endline results. 
We show control means and intervention impacts for primary outcomes in Table \ref{t:itt_fullspec_primary}, and for secondary outcomes of wealth, beneficiary welfare, and cognitive and non-cognitive skills in Tables \ref{t:itt_fullspec_secondary_wealth}, \ref{t:itt_fullspec_secondary_welfare}, and \ref{t:itt_fullspec_secondary_skills} respectively.
These are generally attenuated since those observed at midline, when compared with a control group who have maintained their employment rate and even slightly increased both income and consumption by increasing productive hours (or sacrificing labor) and, most strikingly, stripping productive assets.

Individuals in our control group have an endline employment rate of 50 percent, statistically indistinguishable from the 48 percent employment rate observed at midline. 
Control group incomes are actually 12 percent higher in real terms than they were at midline, and real consumption in the control group is up approximately 47 percent at endline relative to midline.  
These successes in sustaining income and consumption among control-group members have been accompanied by rising work hours and reductions in productive asset stocks.  
Work hours have increased by nearly one hour per week, or approximately 4 percent, implying that earnings per productive hour have risen slightly in the control group between the midline and endline.  
On the other hand, control-group participants have seen marked drops in the value of productive assets they hold:  a loss of approximately 63 percent of the value of productive assets from midline to endline.  
Falls in livestock wealth appear to be a primary contributor to these declines in productive asset values; we also see rising debt among the control group (and, perhaps surprisingly, some increases in savings stocks).  

Against this context, ITT results for primary outcomes in Table \ref{t:itt_fullspec_primary} show proportionally reduced estimates on income, assets, and consumption, with a striking change in the ordering of treatment effects on productive hours, for which HD's impacts have actually grown to make this now the leading intervention.  Aggregate employment effects, combining all types of work, remain essentially unaffected.  
Monthly incomes, which had risen by between 70 and 114 percent among cash and combined cash and training arms at midline, are substantially reduced, with only the Large transfer sustaining a statistically significant, 72 percent increase in monthly income (very similar to that arm's effects at midline).  
Similarly, impacts on household consumption per capita are no longer statistically significant, with point estimates falling to as little as half of their previous value.  Consumption impacts relative to control of 21 percent in GD Middle and of 17 percent in the Combined arm, each of which is about two thirds of its midline impact, have multiple-inference adjusted (``sharpened'') $q$-values just above 10 percent.\footnote{The stars in our tables use the multiple inference-corrected Q-values from \cite{anderson2008multiple} to account for the multiple outcomes and treatments being tested in each table.}  To the extent that consumption is the typical omnibus measure of economic welfare, it may be more appropriate to consider the unadjusted p-values for this outcome, in which case cash arms of the Middle or larger sizes have impacts that are significant at the 95\% level.
Program impacts on productive assets stocks had at midline increased by 154 percent in the HD arm and approximately 400 percent across cash transfer and combined arms.
These differences are now smaller relative to control---with impacts ranging from 93 percent in the HD arm  to as much as 315 percent in the Combined arm---and given the reduced level of productive assets in the control arm, these reflect even smaller absolute differences than they did at midline.\footnote{Because the cash arms have virtually 100\% compliance, the ITT estimated here is also the Treatment on the Treated (ToT).  For the HD arm where the core measure of compliance is 85.6\%, if we are willing to assume that those not participating received no indirect effect of being included in the treatment, then we can back out the ToT by dividing by the compliance rate. The resulting ToT estimate would 17\% larger than the ITT for each variable, with the same significance level.}

In Table \ref{t:employment_breakdown}, we see that even 3.5 years after intervention, the programs studied have induced remarkably stable transitions in sectors of occupation, in spite of the lack of movement in the overall employment rate.\footnote{Note that because this whole table effectively studies a single outcome: ``how are beneficiaries using their time'' we base the stars on the unadjusted p-values, although the sharpened $q$-values are also provided in hard brackets.}   Cash transfers in the Upper, Large, and Combined arms have induced statistically significant levels of entry into non-agricultural microenterprises among between 11 and 17 percent of the population assigned to those arms, with a further 5-6 percent induced to enter into non-agricultural microenterprises in the Large and Combined arm. As in the midline, this appears to be mostly associated with a commensurate movement out of agricultural wage labor, the prevalence of which is also reduced by 8 percentage points by the HD intervention.  Although no longer statistically significant, estimated HD-induced movements into non-agricultural wage labor of 5 percent of the population remain consistent with those found at midline.\footnote{We note that neither treatment induces exit from rural areas altogether; Table \ref{t:res_status} examines whether individuals live in a location classified as non-rural by the Rwandan government, and whether they have moved since baseline.  No significant differences are found; while both treatments weakly increase moving to new villages it is actually \emph{less} likely that treated individuals are located in urban areas.}  Table \ref{t:education_time} examines two related margins, looking at unpaid uses of time (education and domestic work), as well as survey-based measures of the opportunity cost of time, and finds no treatment effects on any of these margins.  Taken together, these findings suggest that cash transfers---particularly those above the Middle value---induced movements into self-employment that came at the cost of own-farm agricultural work and persisted in spite of the pandemic.  On the other hand, while HD's impacts on movements into wage labor did persist, it seems plausible that the more limited productive-asset buffer in those sectors explains the lesser persistence of HD-induced microenterprises over the course of the pandemic.

Similar fading of endline impacts relative to the midline is observed for measures of beneficiary welfare in Table \ref{t:itt_fullspec_secondary_welfare}.  HD impacts on subjective well being fall from 0.19 standard deviations at midline to (statistically insignificant) 0.12 standard deviations at endline, while cash transfer impacts are approximately half of those previously observed---though still significant at 0.29 and 0.39 standard deviations, respectively, for the Middle and Upper transfer values.   We see no impacts on our survey measures of mental health, and we see modest impacts on beneficiary-specific consumption in the Upper and Combined transfer arms.  

As Table \ref{t:itt_fullspec_secondary_wealth} shows, we continue to see signs of persistent wealth effects from cash transfers, though these are generally smaller than at midline and somewhat imprecisely estimated.  Point estimates for net non-land wealth suggest gains in the Upper and Large transfer arms of 70--80 percent relative to control---though statistically insignificant---down from impacts in excess of 110 percent at midline.  There remain positive impacts on household livestock wealth from the Middle, Large, and Combined arms, with the latter delivering the smallest of these impacts at approximately 126 percent over control.  And savings impacts of cash transfers largely persist from midline, e.g., at a 99 percent impact over control in the Middle arm.  The prior estimates of large savings impacts of HD have largely evaporated.  

Finally, we see little evidence of sustained impacts on beneficiaries' cognitive and non-cognitive skills in Table \ref{t:itt_fullspec_secondary_skills}. We see no impacts on aspirations (and endline survey data do not provide the Locus of Control measure from midline).  Measures of business knowledge remain statistically significant in the HD and Combined arms, suggesting some persistent human capital effects, but these are approximately half of their prior magnitudes.

\subsection{ITT Results on Business Outcomes}

We begin our analysis of beneficiary businesses on the extensive margin, examining firm birth and death.  In Table \ref{t:business_survival} we consider the number of newly born firms in midline and in endline in the first two rows. The average control individual created 0.5 new businesses in midline, HD elevates this by 0.22, and all of the cash arms lead this to more than double.  Control individuals created 0.24 new firms on average between midline and endline, and only the GD Large and Combined arms continuing to elevate business creation by midline.   In the remaining rows we then look at the (endogenous) sample of firms that exist at midline, and ask what happens to them at endline as a function of treatment status.  In the third row we see that in the control group 0.24 of the firms have died, this rate is not significantly different for any treatment arm.  In row four we look at the rate of `inoperative' businesses, however, and see that while this is relatively rare in the control group (0.14) the rate of midline businesses becoming inoperative more than doubles in most of the cash arms.  However, because the overall number of businesses was so much larger in the cash arms, they are have more businesses created in midline that remain operative in endline as well.  So, the takeaway from this table is that all of the interventions led to a short-term burst of business creation; while this effect was smaller in the HD arm those businesses proved more durable.  The cash arms created more businesses that survived to endline but also more businesses that become inoperative by endline as well.  So the interventions have powerful effects on the extensive margin.

We turn to the impacts on midline beneficiary businesses in Tables \ref{t:business_outcomes_enterprise_midline}.  Here we see transformative effects of all the interventions.  Beginning again with business ownership in the control group, we see that by midline the average control individual reports operating 0.79 and owning 0.71 businesses (the maximum number reported by control individuals is 5 different enterprises, with 52\% reporting owning any business).  The interventions all drive this number up, with HD increasing owned businesses by 0.14, or 18\% of the baseline mean, and the cash arms having at least three times this effect, with the largest transfers and the Combined arm almost doubling the number of businesses owned. All interventions increase hired labor but particularly draw heavily on the use of household workers, explaining the weak negative effects seen on household-reported enterprises.  HD drives up the number of days worked per month by 2.7 over a base of 9.1, and leads to a large increase monthly profits (treatment effect of \$4.36 per month), but does not change either customers or sales significantly.  The cash arms lead to a doubling of days worked, number of customers per month and sales are more than doubled in the smallest arms and tripled in the Combined arm, and profits are more than doubled everywhere.  The implication of this latter result is that the receipt of \$503 was generating an enterprise profit increase of \$11.32 per month 14 months after receipt of the cash grant.   

The endline beneficiary-reported business impacts are presented in Table  \ref{t:business_outcomes_enterprise_endline}.  Perhaps the most important thing to point out here is the sharp contraction in the overall rate of business ownership in the control group, which falls from 0.79 at midline to 0.4 at endline.  Given that we would expect the control group to be becoming monotonically more economically active over time as they age, this is strong evidence of the fact that Covid-19 has driven a substantial number of the self-employed out of business.  Similarly, control group endline days worked in self-employment fall by 38\%, sales by 53\%, and profits by 24\% relative to the midline.  So there seems to be no doubt that business conditions have worsened overall and the endline impacts need to be interpreted in light of their ability to insulate beneficiaries against this shock. 

As is the case with many of the results presented in this study, the core enterprise treatment effects in the endline represent a fade of about 50\% relative to what we saw at midline.  HD continues to elevate the number of businesses owned by about 0.1, and both days worked and sales are significantly elevated relative to the control.  Unfortunately the HD effects on profits have fallen to about a third of what they were at midline (now \$1.64 per month), and are significant only at the 10\% level.  The cash arms elevate the number of endline businesses by 0.2--0.4, retain substantial impacts on days worked and sales, and continue to significantly elevate profits by amounts ranging from \$3.08 (GD Middle) to \$7.11 per month (GD Large).  While it is impressive to see significant impacts on business profits across the board even 42 months after the GD intervention, these endline impacts represent between 30\%--90\% of the profit impacts at midline, suggesting that all the cash arms are seeing a fade in business profits over time.\footnote{One admittedly heroic way of contextualizing these cash effects is as follows: take the business profit impact of GD Upper 14 months and 42 months after treatment and linearize it from month 1 until it becomes negative (which occurs in month 65) then the sum of the resulting profits is \$463 for an arm that cost \$572 and delivered \$503.  The implication is that the average total improvement in business revenues is not larger than the original transfer, despite the very substantial increase in days worked over the course of this time interval (450 total additional days, using the same linear extrapolation method).  Even taking the total profit effect as return (ignoring the initial cash received) this method suggests a wage rate of around a dollar per day the business is open, similar to the rate observed in the control group.}  Given that the treatment effects on profits and hours worked contract by similar amounts, impact on the effective wage rate appears similar at midline and endline. 

Table \ref{t:beneficiary_endline_gender_int} shows the gender interactions with treatment on endline beneficiary business outcomes.  While the table contains very few significant gender interactions, the signs and the magnitudes of the differential female effects are troubling.  For the key outcomes of sales and profits the female interactions are negative for all treatments, implying that women are benefiting less than men.  While for the cash arms these interaction effects are smaller than the male treatment effect, meaning that women still benefit overall, for HD this is not the case.  Adding together the uninteracted male effect with the female coefficient to get the total effect on women implies that females are getting little endline benefit from HD on customers, sales, or profits compared to women in the control group.  So the modest long-run effects of HD on business outcomes appear to be confined to males (although the difference between men and women is not significant).

These impacts appear to be specific to enterprises owned and operated by the beneficiary, with little impact on businesses of other household members.  We report results for other household-reported enterprises in Tables \ref{t:business_outcomes_hh_midline} and \ref{t:business_outcomes_hh_endline}.  In both rounds, these businesses seem to have been largely untouched by the substantial interventions being directed at youth in these households.  As a starting point we see that control households report operating only 0.06 businesses on average in midline and endline (there is only one control household operating more than one business, so in effect this means that only 6 percent have any business at all).  Treatment of beneficiaries does not lead to any elevation of the probability that there is a household-reported business, and if anything seems to lead to a light \textit{decrease} in the devotion of household labor to the household enterprise.  Similarly, core businesses outcomes such as the number of customers, daily sales, and monthly profits typically show weak negative effects, stronger in the endline than midline.  So the main picture is that there are very few businesses in these households, they are generally unaffected by the presence of the treatments, and to the extent that they are impacted they appear to be suffering from a drawing away of labor.

In summary, then, even the GD Lower arm drives larger enterprise impacts than HD across most outcomes at both midline and endline.  Business conditions have worsened substantially overall between midline and endline, and the impact of the interventions has faded by about half across most outcomes in the 28 months between these two surveys.  These results indicate that enterprise activity is a key conduit for the overall income and consumption impacts seen elsewhere in the study, that both human and physical capital can deliver better livelihoods through self-employment, but neither of these appears to generate a dynamic shift in business outcomes that represents a real pathway out of poverty.

\subsection{Complementarities}

We can more explicitly  test for complementarities by comparing impacts of the Combined arm with the sum of impacts in the arms comprising its constituent parts---HD and the GD Middle transfer arm.  We do so by dropping active-treatment arms not involved in this comparison. We then create indicators for whether the individual received HD or a cash transfer, defining these to take a value of one in the combined arm as well. In Table \ref{t:complementarities_primary} and \ref{t:complementarities_secondary}, we estimate a model that includes these alongside an indicator taking a value of one for individuals assigned to the Combined arm:  the coefficient on the Combined arm indicator therefore directly estimates the extent of complementarities.  

Whereas at midline we had found some evidence of negative complementarities---in particular, on productive hours and subjective well being---we find no such evidence here.  The Combined arm's impacts are statistically indistinguishable from the sum of its HD and GD components.  An even more striking demonstration of this is provided in Tables \ref{t:complementarities_business_midline} and \ref{t:complementarities_business_endline}, which conduct the complementarities tests for the business outcomes where both treatments themselves had strong impacts, particularly at midline.  Somewhat remarkably, neither at midline nor at endline do we find any evidence of complementarity on any outcome, even if we were to use the unadjusted (non FDR-corrected) standard errors.  The study therefore provides no evidence that extra benefits can be unlocked when physical and human capital constraints are relaxed simultaneously.

Given the coincidental fact that the GD Large arm has almost the same cost as the Combined arm, our study provides an alternate ability to think about complementarities.  This is to ask:  given that a youth has already received a cash transfer of approximately the GD Middle amount, is it better then to spend additional resources on that individual by giving her more cash, or by giving the HD training?  This comparison is given in the p-values of the F-tests in column (c) of Table \ref{t:itt_fullspec_primary}, for primary outcomes, and Tables \ref{t:itt_fullspec_secondary_wealth}, \ref{t:itt_fullspec_secondary_welfare}, and \ref{t:itt_fullspec_secondary_skills} for secondary outcomes.  Again, we find no significant differences between these interventions with the exception of business knowledge, showing again that HD is uniquely productive of that outcome.

\subsection{Cost-equivalent benchmark}

A central goal of this project is to allow the comparison of impacts between cash and in-kind programs while holding expenditure per beneficiary constant.  As discussed in \citet{mcintosh2022using}, we do so by estimating a regression-adjusted, cost-equivalent comparison between HD and cash transfers by estimating a model of the form
\begin{equation}\label{eq:CostEquivReprised}
Y_{ihb2}= \delta^T T_{ihb} + \delta^{HD} T_{ibh}^{HD} + \beta X_{ihb0} + \rho Y_{ihb0} + \gamma_1 \tau_c + \mu_b + + \epsilon_{ihb2} 
\end{equation}
for outcome $Y$ of individual $i$ in household $h$, randomization block $b$, and round $2$.  Here, $T_{ihb}$ is an indicator for whether this household was assigned to \emph{any} treatment, and $T_{ibh}^{HD}$ an indicator for assignment to the HD arm in particular, such that the coefficient $\delta^{HD}$ estimates the differential effect of assignment to HD, relative to a cash grant. We include a measure, $\tau_c$, defined as the difference between a given arm's expenditure per beneficiary and that of the HD arm (set equal to zero in both HD and Control); doing so ensures that the coefficient $\delta^{HD}$ estimates HD's differential impacts relative to cash transfers at costs equivalent to HD costs to donors per beneficiary.

At these cost-equivalent levels, we find no statistically significant differences on primary outcomes at endline between HD and cash transfers. Results are presented for primary outcomes in Table \ref{t:costequiv_primary}, which shows the result of this exercise at both midline and endline, and for secondary outcomes at endline in Table  \ref{t:costequiv_secondary}.  Among secondary outcomes, these differences between arms are different only for our measure of business knowledge.  Proportional fading of program impacts across a wide range of outcomes appears to mean not only that it is harder to detect the impacts of individual programs at these more modest impact levels, but, moreover, the resulting attenuation of the differences between arms makes it harder to find statistically significant contrasts between them.  If we willing to interpret differences that appear quantitatively meaningful even if statistically insignificant, HD edges cash in terms of productive hours and debt reduction, while cash has the advantage for outcomes that pertain to asset ownership and consumption.  

We can use this same approach to gauge the relative impacts of cash and training on business outcomes as well.  In Table \ref{t:costequiv_business_midline} we look at midline outcomes.  This table shows that almost across the board cash has been more effective at driving business outcomes.  Owning and operating businesses, days worked and customers per month, daily sales, and monthly profits are all significantly more strongly driven by cash than HD.  Every one of the eight business outcomes studied in this table is also significantly improved by the cost-equivalent cash transfer.  Over the short term then, clearly cash has proven superior to HD in driving business formation and growth.  Table \ref{t:costequiv_business} shows that, as for the primary and secondary outcomes of the study, by the endline the overall effects have faded enough that we can no longer reject equality of cost equivalent impacts for any of the business outcomes.  Nonetheless, the midline results provide powerful evidence that cash `wins' in terms of cost equivalent short-term business operations even up against a program whose specific purpose is to promote self-employment.

Our approach to cost adjustment assumes linearity in the response to spending, an assumption that can be probed by varying the functional form used to control for cost.   Tables \ref{t:costequiv_linearity_primary} and \ref{t:costequiv_linearity_secondary} present the estimated cost-equivalence comparison using a variety of different functional forms to control for cost, for primary and secondary outcomes respectively. In each table, Column 1 present the base linear case from the prior tables.  Column 2 uses a quadratic, and column 3 a third-order polynomial, functional form to control for cost.  Columns 4--7 then serially drop one of the GD transfer amount arms and present the cost equivalence comparison if that arm had not been in the study.  In general the results are quite robust; the significant benefit of HD at building business knowledge is always positive and is significant in 5 of the 7 specifications.

\subsection{Cost-equivalence versus Cost effectiveness}

\begin{figure}[!p]
\caption{{\bf Cost Equivalence versus Cost Effectiveness, Endline}}
\label{f:CE_CEff_Comparison}
\begin{center}
\includegraphics[width=0.95\linewidth]{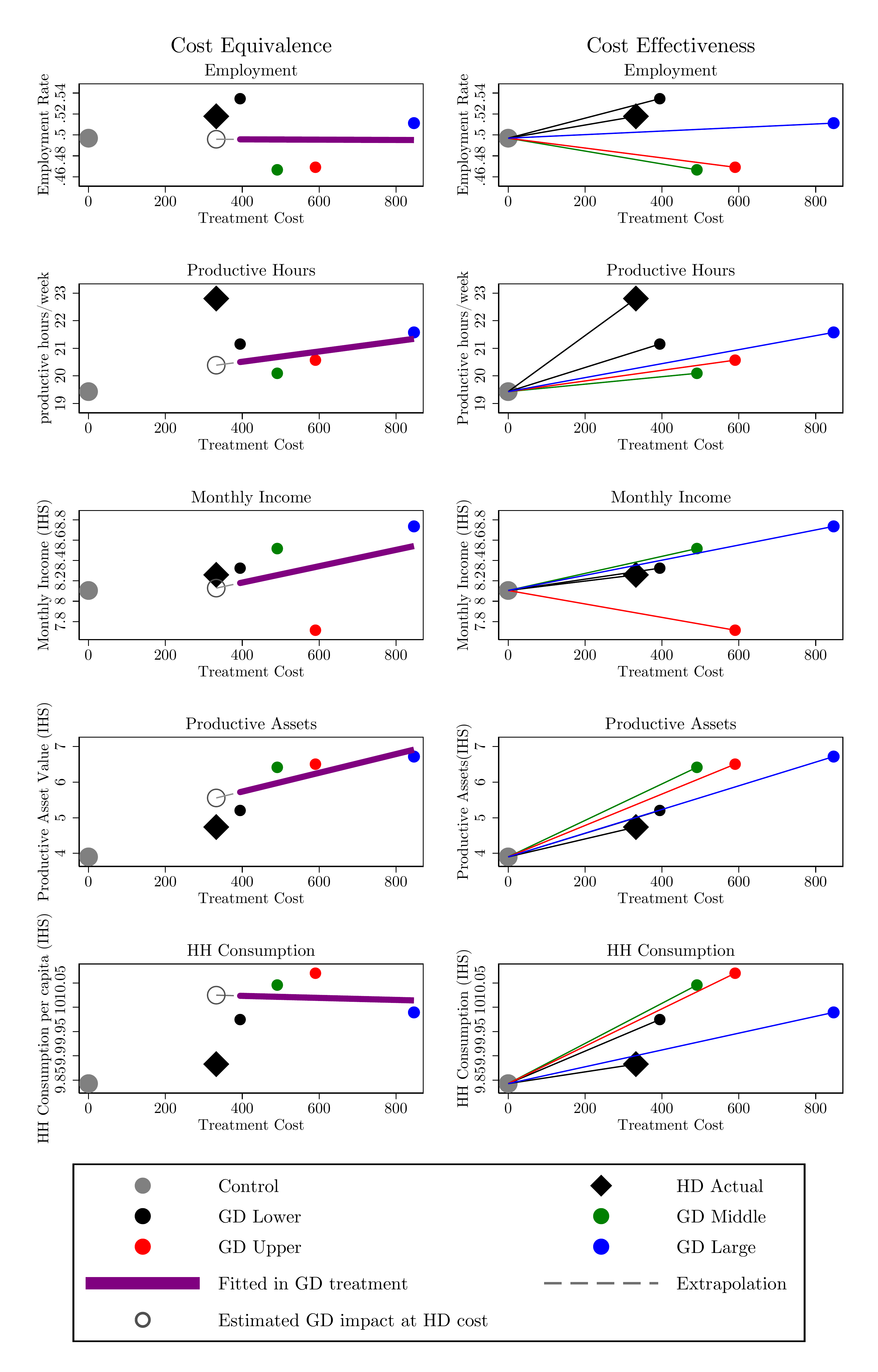}
\end{center}

\end{figure}

This study provides the capacity to make comparisons both across two interventions implemented at (nearly) the same cost, and also to compare across different costs to evaluate differential cost effectiveness per dollar spent.  Tables \ref{t:benefit_cost_ratios_primary} and \ref{t:benefit_cost_ratios_secondary} divide the arm-specific benefits measured in ITT regressions by the cost of each arm in hundreds of dollars, and so give the benefit per amount spent.  The columns to the left of this table then provide p-values on F-tests of the differential cost effectiveness across arms.  As was the case with the cost-effectiveness comparisons, the overall impacts are now sufficiently moderated that none of the benefit/cost ratios are different across arms, with the exception of the business knowledge question.  Figure \ref{f:CE_CEff_Comparison} presents a graphical contrast of the cost equivalence and cost effectiveness approaches to our study results.  Cost equivalence is visualized in the left panels by the vertical difference between the black diamond (HD) and the hollow circle (predicted cash impact at HD cost).  Cost effectiveness is visualized in the right panels by the slope of the line connecting zero with the arm-specific outcome represented in benefit/cost space.  While we have already shown that these differences are not statistically significant, the takeaway from these different approaches emphasizes the superiority of HD at driving productive hours (both in terms of cost equivalence and cost effectiveness), and the cost effectiveness superiority of the middle cash transfers in producing consumption and productive assets.

For a given budget, a comparison in benefit/cost terms is particularly appealing to the extent that results of a cost-effective, inexpensive intervention can plausibly be ``scaled up'' to deliver the same benefit in aggregate welfare terms.  We can empirically address two critical assumptions, the violation of which can constrain that thought exercise.  The first of these is \emph{heterogeneity}: if there exists a large sample of similarly eligible individuals for whom the same benefits can be achieved, then the less expensive intervention can always be ``scaled up'' in aggregate welfare terms on the extensive margin by simply applying it to more people. The presence of meaningful heterogeneity suggests the need for care in identifying a target population for scale-up. The second relates to the nature of \emph{spillover effects} and the extent to which impacts estimated on the target beneficiaries correctly reflect the full welfare benefits of the program at a population level.  In the presence of spillover effects, scale-ups of interventions that operated within a fixed geographical space might deliver different results than those found in less intensive interventions.  

To examine \emph{heterogeneity}, we\ use interaction analysis to examine the pre-specified dimensions over which we anticipated the study might have differential effects.  These are gender (Table \ref{t:heterogeneity_female}), age (Table \ref{t:heterogeneity_older}), baseline consumption (Table \ref{t:heterogeneity_consump}), and baseline local employment rates (Table \ref{t:heterogeneity_employ}).  None of these tables surfaces any meaningful evidence of heterogeneity.  One possible explanation of this result is that our study ended up with a relatively narrow set of targeting criteria (youth who were qualified for and interested in Huguka Dukore, while being poor enough to qualify for the use of Give Directly transfers), thereby limiting the overall diversity within our sample.  The conclusion is that both of the interventions studied are having consistent effects and retargeting within this group would not substantially improve overall program effectiveness on primary outcomes.

To examine \emph{spillovers}, we exploit experimental variation in the shares of HD, GD Main, and GD Huge recipients in a given village in order to test for the presence of local spillovers.  We estimate two types of models of spillovers:  a \emph{levels spillovers} specification that allows for neighbors' treatments to affect one's own outcomes in levels, and a \emph{general interference} model that allows not only for these levels effects, but also for the share of individuals in each arm to modify the impacts of each treatment. Across these models, we find only very limited evidence of program spillovers at follow-up.  As Table \ref{t:saturation_levels} shows, in the levels spillover model we see some evidence of a negative spillover from HD saturation levels onto peer consumption.  However, there is limited additional support for this estimate in the richer interference model for consumption outcomes presented in Table \ref{t:interference_hh_month_consumption_pc},  where point estimates for spillovers from HD treatment onto control and cash-transfer arms are negative but insignificant. If anything, the full saturation model suggests a statistically significant, negative spillover effect of the GD Main treatments onto consumption levels in Control.  We see no further evidence of interference on other outcomes in the analysis of spillovers to employment (\ref{t:interference_bn_employed}), income (\ref{t:interference_bn_monthly_income}), productive hours (\ref{t:interference_bn_productive_hrs}), or asset values (\ref{t:interference_bn_tot_prod_assetval}).  Hence the within-village variation in treatment intensity reveals very little evidence of contamination in the study.

The conclusion of these two empirical extensions is that the estimates provided here are a) broadly applicable to the entire target population and b) do not generate strong externalities.  Hence the linearization assumption underlying the standard cost effectiveness approach may be justified in the sense that any of the less expensive interventions studied here can be scaled up in terms of total welfare benefits simply by applying them to more similar people.

\subsection{Contextualizing the Covid-19 shock}

\begin{figure}[!hbtp]
\caption{{\bf Incidence of Covid-19 Shocks, by Treatment Arm}}
\label{f:shocks_by_tx}
\centering
\includegraphics[width=0.75\textwidth]{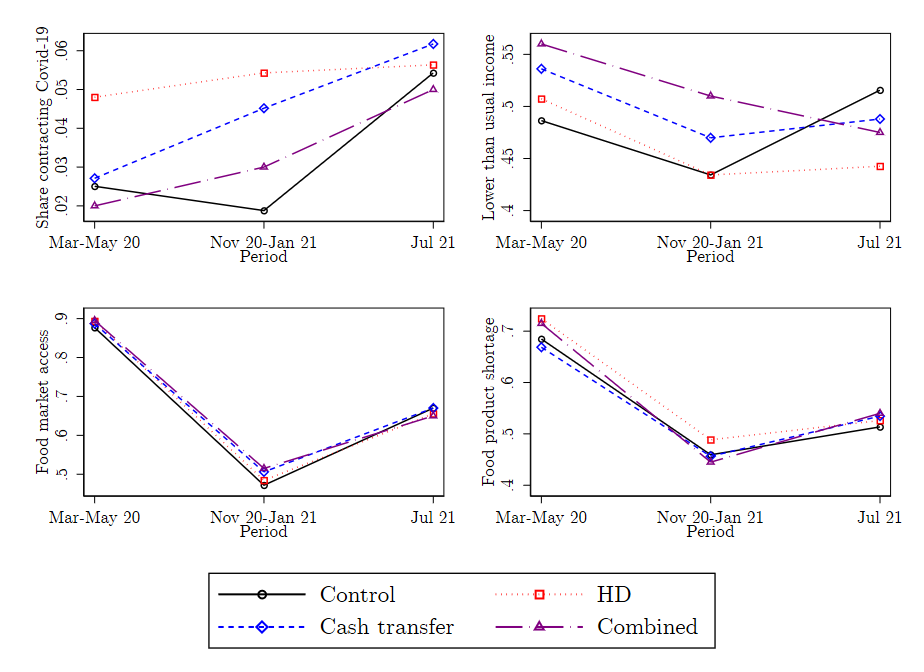}
\end{figure}

To understand the incidence of Covid-related shocks, we asked study participants at endline about their experiences during three key and salient periods:  first from the beginning of the nationwide lockdown of March 2020 until the beginning of the genocide memorial period in May, 2020; second, in a period of relative normalcy, during which children were allowed to return to school, running from November 2020 through January 2021; and third, during a later lockdown in July of 2021. In Figure \ref{f:shocks_by_tx}, we document the incidence of four measures of shocks during this period.\footnote{We focus on those that are plausibly exogenous to assigned treatments, excluding those (e.g., asset stripping) that are more likely to be driven by treatment-induced accumulation of wealth or economic opportunity.  We report the incidence of actual disease incidence, self-reported income loss, and experiences of food-market closures and food product shortages.}   Variation in these outcomes across arms is of modest magnitude and statistically imprecise.  Reports of direct experience of Covid illness are relatively low in most periods, though increasing over time.   We see markedly greater measures of induced economic hardship in the first of the reference periods, with some resurgence of these shocks in the later lockdown period. In particular, a substantial portion of the sample reports income losses, particularly in the first lockdown period.  And the consequences of this lockdown are also visible in reported experiences of access to food:  both reductions in market access and shortages of specific products are widespread, especially in the initial lockdown period. 

\begin{figure}[!hbtp]
\caption{Evolution of control-group primary outcomes across survey rounds}
\label{f:control_primaries_round}
\centering 
\includegraphics[width=0.75\textwidth]{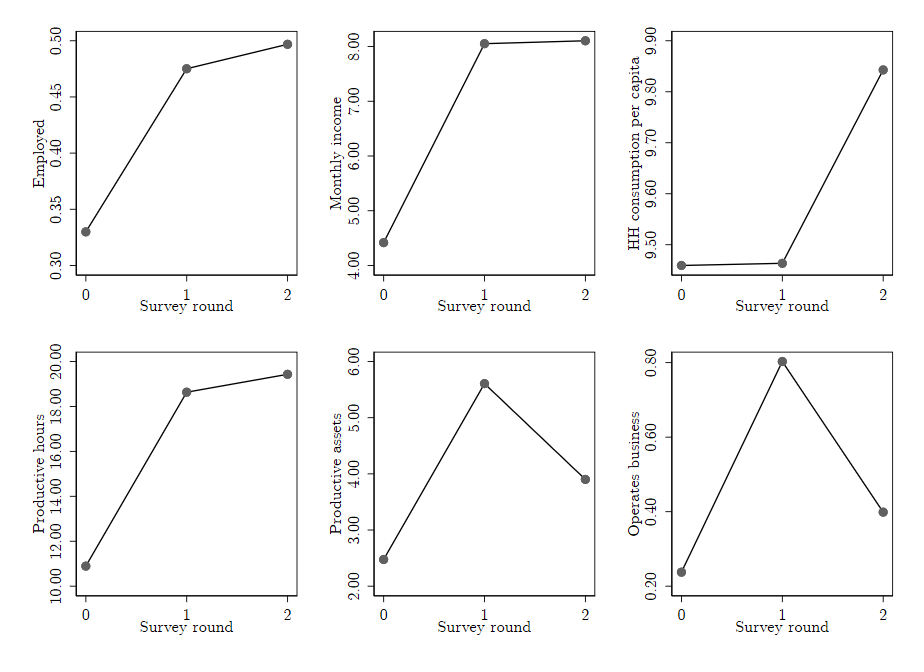}

\floatfoot{
	\begin{footnotesize}
	Notes: Endline outcomes deflated to nominal midline Rwanda francs.
	\end{footnotesize}
}
\end{figure}

These shocks are borne out in stalled improvements in income-generating activities in the control group.  This is documented in Figure \ref{f:control_primaries_round}.  The growth trajectory in employment status, productive hours, and monthly income comes to a near-complete stop between the midline and endline, however, as control members experience the economic consequences of the Covid-19 pandemic.  Control-group members were able to protect consumption at endline, but potentially at the cost of a stripping of productive assets.\footnote{Consumption figures are deflated to midline prices, so this does not reflect an inflationary effect.} Control-group households lose more than half of the value of their assets at midline and the rate of business operation contracts markedly.  Taken together, these findings suggest that individuals in the control arm reduced or sold off the businesses that they had launched by midline; in doing so, they did not become unemployed, but rather switched back to a focus on agriculture, with some surplus generated by asset sales that drove a rise in consumption.

Our ability to distinguish the consequences of Covid-19 shocks hinges on the existence of measurable, cross-sectional variation in exposure to exogenous measures of shocks.  Unfortunately,  the large majority of shock incidence is time-series:   the geographic sector intra-class correlation is less than 0.01 across all four of the ``exogenous'' shock measures highlighted above.  A separate way of making the same point is shown in Table \ref{t:tx_protect}, where we follow the ``endogenous stratification'' approach suggested by \citet{AbaChiWes18restat} to predict exposure to Covid-19 shocks using a set of baseline covariates.  The ``index'' variable refers to the predicted value of that outcome in the control group.   Overall, we see very limited predictive power of this index on endline outcomes.  Perhaps unsurprisingly given this, there is also little evidence of heterogeneity in treatment effects across levels of the predictive index, with the exception of the GD Large treatment arm's impacts on employment and consumption.\footnote{If attenuation in treatment effects since midline were attributable to the observed shock measures, we would expect coefficients on the interaction between predicted endline outcomes and treatments to be positive:  this would indicate that income losses are more severe among the treated.}  For those outcomes, results are suggestive of particularly strong deterioration in outcomes among those GD-Large recipients who experienced negative shocks.  

Finally, in Table \ref{t:control_covid_outcomes} we use a broader set of measures describing the incidence of shocks (including those more directly endogenous to behavior), and examine how these correlate with the key outcomes of the study.  This analysis shows very clearly that the reported incidence of economic shocks from Covid-19 is actually \textit{higher} in households that were more deeply engaged in non-farm entrepreneurship.  Strong positive coefficients are observed between the economic blow from Covid-19 and measures of economic activity, including employment, productive hours, income, and productive assets (both in levels and changes).  The implication is that those who were most exposed to the small business sector were hit hardest by Covid-19.

In sum, then three pieces of evidence suggest that the Covid-19 shock hurt those who had ``stuck their neck out'' by engaging in entrepreneurship, which is of course precisely what these interventions intended to do.  First, the absolute fall in measures of entrepreneurship during the Covid-19 era are larger for the treatment groups than the control.\footnote{Midline to endline changes in business ownership rates are -0.39 in the control, -0.43 for HD, and -0.63 for GD large.  IHS productive asset changes are -1.6, -2.3, and -2.6 respectively.}  Second,  Table \ref{t:tx_protect} shows that the kinds of individuals who were most likely to be hit by Covid-19 shocks in many cases have somewhat larger income fluctuations if treated than if in the control.  Finally,  table \ref{t:control_covid_outcomes} illustrates that within the control group reported incidence of shocks to income were actually \textit{higher} among those with higher employment rates, income, and productive assets.  So the overall takeaway is that this period was particularly difficult for people who were doing the things that the treatments intended to induce (running their own businesses rather than working on the farm).  Consequently, the presence of the Covid-19 shock is likely to have attenuated the enterprise-driven income effects observed in this study (although the Covid lockdowns likely hit the self-employed sector harder than agriculture in a way that would not be seen from other shocks, such as drought or food price changes).  This illustrates the role that shocks can play in eroding the gains that transfers can enable.  While it might appear that human capital interventions would prove robust to such shocks, in this context we find the proportional fade to be almost perfectly symmetric.

\section{Aggregating Impacts over Time}\label{s:aggregation}
%!TEX root=./ms.tex

We conclude our empirical analysis with an exercise intended to sum up all of the financial flows observed through both follow-up rounds of the survey to estimate the total effect of the cash intervention on household financial flows.  Aggregating impacts across financial outcomes and over time provides one approach to summarizing the relative impact of each intervention studied.

This exercise us requires us to impute values of financial flows between study measurement points.  Given that our study asks beneficiaries about wealth, transfers, income, and consumption with different recall periods as recommended by the survey literature, and given that we do not have survey data explicitly asking about these flows in all months from baseline to endline, aggregating the total flow of resources during the course of the study necessarily requires making some strong assumptions about how these flows change over time. The extent to which our measures of inflows, expenditures, and asset value changes \emph{balance}---that is, the extent to which we can account for all impacts on beneficiary income in either expenditure or asset accumulation---will provide one indication of the accuracy of these assumptions.

The simplest values to account for are stocks; the survey directly asks for the current value of things such as productive assets, livestock, savings, and debt, and so the total effect of the treatment on these values at a moment in time is fully reflected by differences in stock value.  An intermediate case are irregular flows that asked about in the survey `over the past year', which include transfers made to and received from other households.  It is standard to ask these questions at a longer time frequency since these flows tend to be large and irregular, meaning that short recall windows become very noisy.  Then we have variables that are measured as short-term flows, of which the most important are consumption and income.  Here we ask questions either over a month recall window (durables) or a week (non-durables) and so can aggregate consumption and income to monthly levels over the month prior to the survey.  For the annual and monthly flow variables, to calculate total impacts over any period of time, an assumption must be made about how these flows change during the course of the study.

At midline we conducted a cash accounting exercise covering the 12 months since treatment by summing the stock values, the annual flows, and then multiplying the monthly flows times 12 which implicitly assumes that the impacts seen a year out had been exactly sustained during the course of that year.  We now attempt to repeat this exercise for a survey conducted 40 months after baseline and 28 months after midline.  To aggregate values at this point we assume the flows followed a step function, taking their midline values up through the midline and their endline values between the midline and the endline.  Total stock values are simply the endline treatment effects on stocks.  Annual flow values are then the midline treatment effect plus 2.33 times the endline annual treatment effects (reflecting the ratio of the period between midline and endline to the duration of the one-year recall period), and the monthly flow values are 12 times the midline treatment effect plus 28 times the endline treatment effect.  The estimates arrived at through this step-function assumption are conservative relative to the other obvious assumption which would involve a linear interpolation of the outcomes through the midline and endline outcomes for the duration of treatment.  The ingredients for this exercise, then, can be seen in the midline and endline monthly flow income and consumption ITTs in Table \ref{t:itt_fullspec_primary}, the midline and endline annual flow impacts on intra-household transfers shown in Table \ref{t:itt_transfers}, the endline treatment effect on stock variables in Table \ref{t:itt_fullspec_secondary_wealth}.

The results of this exercise are presented in Table \ref{t:cash_accounting}.  The top row of the table provides the cash amount received in each arm.  A starting point is to observe that, relative to impacts observed at midline, 50--85\% of the original transfer is still present in the total stock value of assets, depending on arm.  So the incremental wealth value of the transfer has not been spent down, even in a period in which the control group was spending down assets on the whole.  If the remainder had been simply spent without ever producing any income, we would then expect to see the total inflow equal the transfer amount, and the total consumption equal the difference between the original transfer and the current stock impact.  Instead, we see that the outflows exceed the difference between transfers and stock values, but an amount ranging from about \$140 in the lower and large arms to \$430 in the middle arm.  So what then is the source of this extra money?  The answer is clear in the total inflows; in every case the total inflows greatly exceed the value of the transfer.  The ratio of the total inflows to the transfer is a simple measure of the `multiplier' effect of the cash; this ratio varies from 1.3 in the upper arm to 2.5 in the middle arm, showing that in all arms the transfers were put to work to create additional income.  In all arms the additional income generated (over and above the transfer) is similar to or larger than the spending not accounted for by the draw-down in asset value (original transfer minus current stock impacts).\footnote{The last row `Survey share accounted' gives the fraction of outflows expected as a function of changes in net inflows that we are able to capture.  If all measures are complete, this is an accounting identity, so the ratio of outflows to inflows provides a measure of survey quality.}  

This approach to aggregating financial impacts across the full duration of the study also allows an omnibus test of the statistical significance of cumulative impacts relative to control.  To undertake such a test, we undertake a randomization inference exercise:  we permute treatments to provide alternative randomizations consistent with our block-randomized allocation.  To test the significance of individual arms, we permute (within blocks) the assignment between those arms and control; to test the significance of comparisons between arms, we permute assignment of the relevant pairwise combinations of arms, again within blocks.  Randomization inference provides a distribution of the total income, total expenditure, and total final stock values.  We compare the realized differences between arms to these permutation distributions to obtain a randomization inference $p$-value. 

Consistent with ITT results, this exercise in Table \ref{t:cash_accounting} confirms the statistical significance of cumulative expenditure and stock-value effects of all cash-transfer arms, relative to control.  Cash-transfer effects on cumulative income are significant for the Middle and Large transfer values.  The Combined arm has significant effects on cumulative values of each of income, expenditure, and asset values.  HD alone has statistically insignificant cumulative effects on income and expenditure, and just misses significance at the ten percent level for its impacts on final stock values ($p=0.11$).  

Clearly, on average, the cash transfers have been put to work to drive substantial additional income, enabling outflows to increase by a total of 65 to 120\% of the transfer amount while leaving the majority of the transfer intact after almost four years in asset values.  At the same time, the fact that arms that more than double the value of productive assets lead only to a 20\% increase in consumption at endline suggests that the return on these assets is low, and the labor devoted to operating them is receiving an effective wage rate similar to counterfactual uses of time.  So these transfers are used by beneficiaries in a careful, productive way with an eye to the long term, but the opportunities for transformative enterprise-driven growth appear limited.

%--------------------------------------------------------------------------%
\section{Conclusions}\label{s:conclusions}
%!TEX root=./ms.tex

This study provides a unique window on the comparative effects of efforts to help disadvantaged youth climb the economic ladder.  Given clean experimental variation, high treatment compliance rates, excellent survey tracking for more than three years, and little apparent contamination from spillovers, it is a very straightforward environment in which to view the relative benefits of cash versus in-kind programming in the medium term.  Divergent trajectories are apparent, with workforce training weakly increasing paid work and strongly elevating the chance of full-time employment, and cash transfers enabling self-employment and engendering the creation of profitable businesses that survive for years.   Both interventions induce quantitatively large improvements in productive assets. The effect of HD on productive hours, and the effect of cash on productive assets, remains relatively constant across midline and endline, suggesting that these impacts may prove durable in the long term.  Meaningful changes in the lives of beneficiaries are visible years after the interventions.

On the other hand, the general pattern is that the benefits that are seen in this 3.5-year endline are about a half of what was observed in the midline one year after treatment.  This suggests that---while there were significant effects on economic well-being during the study duration---over the longer term, beneficiaries are on a slow slide back towards the outcomes that they would have achieved in the absence of the programs.  Roughly half of the new enterprises started at midline were no longer operative at endline, and the critical final outcomes of income and consumption are no longer improved relative to the control.  So the takeaway is that while HD beneficiaries are indeed working more, and cash beneficiaries are operating more businesses with larger productive assets,  ultimately the economic returns of these activities to the youth may not be higher than the own-farm agriculture or agricultural wage labor that they would typically have been doing otherwise.  This suggests more systemic problems with the nature of the markets in which these youth work.  Low returns to skill in local labor markets will limit what can be gained through training programs, and weak demand will constrain the potential of a self-employment led exit from poverty.  So the fact that such substantial and expensive interventions do not transform the lives of disadvantaged youth refocuses attention on the macro constraints to growth that limit the ability of individuals to climb out of poverty.

A critical contextual factor for this study is that the period between the midline and endline includes the Covid-19 era.  While many programs that find diminishing long-term impacts use the language of ``catch-up by the control'' \citep{blattman2018long}, in this case the control group has seen a 30\% drop in productive asset value between midline and endline, and the share of controls operating businesses dropped from 79\% to 40\%.  Since both interventions induced beneficiaries to go into business and put assets at risk, the treatment groups were more exposed to the Covid-19 shock at the same time as they had more wealth to protect themselves against it.  While the treatment groups lost more productive assets between midline and endline than the control during Covid, they nonetheless retained more by endline as well.  While more treatment-induced businesses became inoperative during Covid, because so many had been created initially, still more survived at the end than in the control.  Evidence in this study suggests that it may be the limited ability of interventions to shield beneficiaries from the consequences of such shocks---as much as the rising tide of economic outcomes among those not receiving such benefits---that drives the diminution of program impacts in the long term.  While the specifics of the Covid-19 shock were certainly unique, uncertainty and disruption are an unfortunate fact of life for entrepreneurs in developing economies.  Hence, this mixture of treatment and resiliency impacts may provide a realistic picture of the extent to which these interventions are able to deliver longer-term benefits that persist through good times and bad.  

Over the longer term we can focus our emphasis on the core outcomes of income, consumption, and subjective well-being that exemplify the ultimate impacts on economic welfare, rather than things like business assets or business knowledge that are merely instrumental to long-term welfare.  In the endline none of these outcomes have significant cost-equivalent differences, but for household and individual consumption, subjective well-being, and business income the point estimates all suggest an advantage for cash.  Only overall income is somewhat higher at endline for HD, which hints at the possibility of more persistent income effects from the human-capital investment of its training.  In the midline every one of these outcomes is significantly better for cash at cost-equivalent levels.  Integrating these two snapshots in time over the entire duration of the study, then, it seems relatively clear that cash has done a better job of moving ultimate welfare outcomes at cost-equivalent levels than HD.  These results suggest that investments seeking create economic well-being over the short- to medium-term will do well to incorporate cash as at least a part of their programming.

%--------------------------------------------------------------------------%
%--  BIBLIOGRAPHY --% 
\clearpage
\bibliographystyle{aer}
\bibliography{HD2_references}
%------------------------------------------------------------------------------%
\cleardoublepage

%--------------------------------------------------------------------------%
%       Tables and Figures                                                 %
%--------------------------------------------------------------------------%

%!TEX root=./ms.tex

\begin{landscape}
\section*{Tables and Figures}
\vspace*{\fill}
\begin{table}[!hb]
\caption{Results of Costing Exercise}
\label{t:costing}
\begin{center}
\begin{tabular}{l c c c c c c}
\toprule %
Treatment             Arm:	&	Ex Ante Cost	&	Value received 	&	Ex Post Cost	&	Fraction  operating cost 	&	Compliance Rate 	&	Cost per study household	\\
Huguka Dukore	&	\$464.25	&	\$153.47	&	\$388.32	&	60.5\%	&	85.6\%	&	\$332.27	\\
GD lower	&	\$377.03	&	\$317.16	&	\$394.39	&	19.6\%	&	100\%	&	\$394.39	\\
GD mid	&	\$464.25	&	\$410.65	&	\$493.96	&	16.9\%	&	99.4\%	&	\$490.99	\\
GD upper	&	\$571.74	&	\$502.96	&	\$590.41	&	14.8\%	&	100\%	&	\$590.41	\\
GD large	&	\$828.47	&	\$750.3	&	\$846.71	&	11.3\%	&	100\%	&	\$846.71	\\
Combined	&	\$928.5	&	\$561.11	&	\$885.64	&	36.3\%	&	89.6\%(HD), 100\%(GD)	&	\$840.20	\\

\bottomrule %
\end{tabular}
\end{center}
\vskip-4ex  
\floatfoot{ 
\begin{footnotesize}
Note:  The first column shows the ex-ante costing data on which study was designed; the core number is the HD cost around which the GD actual transfer amounts in column 2 were designed.  Column 3 shows the results of the ex post costing exercise.  Column 4 provides the share of spending that did not reach the beneficiaries either in cash or in direct training and materials costs.  Column 5 shows the compliance rates, and since all costs are averted for non-compliers then the final column shows the final cost per study subject for each arm that are the basis of the cost-equivalent comparisons.
\end{footnotesize}
}
\end{table}
\vspace*{\fill}

\end{landscape}

\begin{landscape}

\begin{table}[h]\centering
\caption{{\bf Simple ITT, Primary Outcomes}.}
     \begin{footnotesize}
 \label{t:itt_fullspec_primary}

\resizebox{.95\textwidth}{!}{ 
\centering   
\begin{tabular}{l *{6}{S} ScSSSSS}
\toprule
 & & \multicolumn{4}{c}{GiveDirectly} & & \multicolumn{1}{c}{\raisebox{-1ex}[0pt]{Control}} & & & \multicolumn{3}{c}{$ p$-values} \\ 
\cmidrule(lr){3-6} \cmidrule(lr){11-13}
\multicolumn{1}{c}{\text{ }} & \multicolumn{1}{c}{\text{HD}} & \multicolumn{1}{c}{\text{Lower}} & \multicolumn{1}{c}{\text{Middle}} & \multicolumn{1}{c}{\text{Upper}} & \multicolumn{1}{c}{\text{Large}} & \multicolumn{1}{c}{\text{Combined}} & \multicolumn{1}{c}{\text{Mean}} & \multicolumn{1}{c}{\text{Obs.}} & \multicolumn{1}{c}{\text{$ R^2$}} & \multicolumn{1}{c}{\text{(a)}} & \multicolumn{1}{c}{\text{(b)}} & \multicolumn{1}{c}{\text{(c)}}\\
\midrule
\multicolumn{13}{l}{\emph{Panel A. Midline}}  \\ 
\addlinespace[1ex] \multirow[t]{ 3 }{0.15\textwidth}{Employed }  &      0.02 &      0.03 &      0.03 &      0.01 &      0.02 &      0.01 &      0.47 &      1770 &      0.16 &      0.98 &      0.63 &      0.90 \\ 
 & (0.03)  & (0.05)  & (0.05)  & (0.05)  & (0.05)  & (0.04)  \\  & [    0.24]  & [    0.26]  & [    0.24]  & [    0.39]  & [    0.38]  & [    0.39]  \\ \addlinespace[1ex] 
\multirow[t]{ 3 }{0.15\textwidth}{Productive hours }  &      2.85\ensuremath{^{**}} &      2.53 &      5.05\ensuremath{^{**}} &      3.99\ensuremath{^{*}} &      1.70 &      2.67 &     18.42 &      1770 &      0.18 &      0.72 &      0.43 &      0.69 \\ 
 & (1.52)  & (2.22)  & (2.21)  & (2.50)  & (2.05)  & (2.01)  \\
\bottomrule
\end{tabular}
}
\floatfoot{
Note:  The six columns of the table provide the estimate on dummy variables for each of the treatment arms, compared to the control group.  The five primary outcomes are in rows.  Regressions include but do not report the lagged dependent variable, fixed effects for randomization blocks, and a set of LASSO-selected baseline covariates. Standard errors (in parentheses) are clustered at the household level to reflect the design effect, and $p$-values corrected for False Discovery Rates across all the outcomes in the table are presented in brackets.  Stars on coefficient estimates are derived from the FDR-corrected $p$-values, *=10\%, **=5\%, and ***=1\% significance. Reported $p$-values in final three columns derived from $F$-tests of hypotheses that cost-benefit ratios are equal between: (a) GD Lower and HD; (b) GD Lower and GD Large; and (c) GD Large and Combined treatments. Employed is a dummy variable for spending more than 10 hours per week working for a wage or as primary operator of a microenterprise.  Productive hours are measured over prior 7 days in all activities other than own-farm agriculture.  Monthly income, productive assets, and household consumption are winsorized at 1\% and 99\% and analyzed in Inverse Hyperbolic Sine, meaning that treatment effects can be interpreted as percent changes.
}
\end{footnotesize}
\end{table}

\begin{table}[h]\centering
\caption{{\bf Simple ITT, Secondary Outcomes: Wealth}.}
     \begin{footnotesize}
 \label{t:itt_fullspec_secondary_wealth}     
 \begin{tabular}{l *{6}{S} ScSSSSS}
\toprule
 & & \multicolumn{4}{c}{GiveDirectly} & & \multicolumn{1}{c}{\raisebox{-1ex}[0pt]{Control}} & & & \multicolumn{3}{c}{$ p$-values} \\ 
\cmidrule(lr){3-6} \cmidrule(lr){11-13}
\multicolumn{1}{c}{\text{ }} & \multicolumn{1}{c}{\text{HD}} & \multicolumn{1}{c}{\text{Lower}} & \multicolumn{1}{c}{\text{Middle}} & \multicolumn{1}{c}{\text{Upper}} & \multicolumn{1}{c}{\text{Large}} & \multicolumn{1}{c}{\text{Combined}} & \multicolumn{1}{c}{\text{Mean}} & \multicolumn{1}{c}{\text{Obs.}} & \multicolumn{1}{c}{\text{$ R^2$}} & \multicolumn{1}{c}{\text{(a)}} & \multicolumn{1}{c}{\text{(b)}} & \multicolumn{1}{c}{\text{(c)}}\\
\midrule
\multicolumn{13}{l}{\emph{Panel A. Midline}}  \\ 
\addlinespace[1ex] \multirow[t]{ 3 }{0.15\textwidth}{HH net non-land wealth }  &     -0.28 &      0.16 &      1.04\ensuremath{^{**}} &      1.26\ensuremath{^{***}} &      0.98\ensuremath{^{**}} &      0.83\ensuremath{^{**}} &     11.41 &      1770 &      0.21 &      0.40 &      0.56 &      0.78 \\ 
 & (0.37)  & (0.53)  & (0.43)  & (0.40)  & (0.40)  & (0.46)  \\  & [    0.28]  & [    0.35]  & [    0.01]  & [    0.00]  & [    0.01]  & [    0.05]  \\ \addlinespace[1ex] 
\multirow[t]{ 3 }{0.15\textwidth}{HH livestock wealth }  &     -0.04 &      1.63\ensuremath{^{***}} &      1.90\ensuremath{^{***}} &      2.66\ensuremath{^{***}} &      2.23\ensuremath{^{***}} &      2.22\ensuremath{^{***}} &      7.88 &      1770 &      0.26 &      0.00 &      0.21 &      0.98 \\ 
 & (0.36)  & (0.49)  & (0.50)  & (0.43)  & (0.45)  & (0.45)  \\
\bottomrule
\end{tabular}
\floatfoot{
    Notes:   Regressions include but do not report the lagged dependent variable, fixed effects for randomization blocks, and a set of LASSO-selected baseline covariates.  Standard errors are (in soft brackets) are clustered at the household level to reflect the design effect, and $p$-values corrected for False Discovery Rates across all the outcomes in the table are presented in hard brackets.  Stars on coefficient estimates are derived from the FDR-corrected $p$-values, *=10\%, **=5\%, and ***=1\% significance.  Reported $p$-values in final three columns derived from $F$-tests of hypotheses that cost-benefit ratios are equal between: (a) GD Lower and HD; (b) GD Lower and GD Large; and (c) GD Large and Combined treatments.
}
\end{footnotesize}
\end{table}

\begin{table}[h]\centering
\caption{{\bf ITT Employment Breakdown}.}
     \begin{footnotesize}
 \label{t:employment_breakdown}
 \begin{tabular}{l *{6}{S} ScSS}
\toprule
 & & \multicolumn{4}{c}{GiveDirectly} & & \multicolumn{1}{c}{\raisebox{-1ex}[0pt]{Control}} \\ 
\cmidrule(lr){3-6}
\multicolumn{1}{c}{\text{ }} & \multicolumn{1}{c}{\text{HD}} & \multicolumn{1}{c}{\text{Lower}} & \multicolumn{1}{c}{\text{Middle}} & \multicolumn{1}{c}{\text{Upper}} & \multicolumn{1}{c}{\text{Large}} & \multicolumn{1}{c}{\text{Combined}} & \multicolumn{1}{c}{\text{Mean}} & \multicolumn{1}{c}{\text{Obs.}} & \multicolumn{1}{c}{\text{$ R^2$}} & \multicolumn{1}{c}{\text{$ p$-value}}\\
\midrule
\multicolumn{11}{l}{\emph{Panel A. Employment composition}}  \\ 
\addlinespace[1ex] \multirow[t]{ 3 }{0.2\textwidth}{Non-agricultural microenterprise }  &      0.05\ensuremath{^{*}} &      0.07\ensuremath{^{*}} &      0.07\ensuremath{^{*}} &      0.11\ensuremath{^{***}} &      0.17\ensuremath{^{***}} &      0.14\ensuremath{^{***}} &      0.16 &      1822 &      0.09 &      0.00 \\ 
 & (0.03)  & (0.04)  & (0.04)  & (0.04)  & (0.04)  & (0.04)  \\  & [    0.16]  & [    0.16]  & [    0.17]  & [    0.03]  & [    0.00]  & [    0.00]  \\ \addlinespace[1ex] 
\multirow[t]{ 3 }{0.2\textwidth}{Other microenterprise or self-employment }  &      0.02 &      0.04\ensuremath{^{*}} &      0.04 &      0.01 &      0.05\ensuremath{^{**}} &      0.06\ensuremath{^{**}} &      0.04 &      1822 &      0.06 &      0.08 \\ 
 & (0.01)  & (0.03)  & (0.02)  & (0.02)  & (0.02)  & (0.02)  \\
\bottomrule
\end{tabular}
\floatfoot{
        Notes: Panel A presents impacts on indicators for employment of any hours in the corresponding activity type in the preceding week.  Panel B presents impacts on an indicator for overall employment, using the reported threshold for minimum hours. Regressions include but do not report an indicator for lagged employment status, fixed effects for randomization blocks, and a set of LASSO-selected baseline covariates. Standard errors are (in soft brackets) are clustered at the household level to reflect the design effect, and $p$-values corrected for False Discovery Rates across outcomes in each panel are presented in hard brackets.  Stars on coefficient estimates are derived from unadjusted standard errors, *=10\%, **=5\%, and ***=1\% significance.
}
\end{footnotesize}
\end{table}

\end{landscape}

\cleardoublepage

\begin{table}[h]\centering
\caption{{\bf Cost Equivalent Analysis, Primary Outcomes}.}
     \begin{footnotesize}
 \label{t:costequiv_primary}
\begin{tabular}{l *{3}{S} ScS}
\toprule
\multicolumn{1}{c}{\text{ }} & \multicolumn{1}{c}{\text{\makecell[b]{Differential impact \\ of HD}}} & \multicolumn{1}{c}{\text{\makecell[b]{Cost-equivalent\\GD impact}}} & \multicolumn{1}{c}{\text{\makecell[b]{Transfer\\Value}}} & \multicolumn{1}{c}{\text{\makecell[b]{Control\\Mean}}} & \multicolumn{1}{c}{\text{Obs.}} & \multicolumn{1}{c}{\text{$ R^2$}}\\
\midrule
\multicolumn{7}{l}{\emph{Panel A. Midline}}  \\ 
\addlinespace[1ex] \multirow[t]{ 3 }{0.2\textwidth}{Employed }  &     -0.00 &      0.03 &     -0.00 &      0.47 &      1578 &      0.17 \\ 
 & (0.04)  & (0.04)  & (0.01)  \\  & [    0.75]  & [    0.49]  & [    0.69]  \\ \addlinespace[1ex] 
\multirow[t]{ 3 }{0.2\textwidth}{Productive hours }  &     -1.26 &      4.16\ensuremath{^{*}} &     -0.37 &     18.42 &      1578 &      0.19 \\ 
 & (2.08)  & (2.02)  & (0.54)  \\  & [    0.49]  & [    0.07]  & [    0.49]  \\
\bottomrule
\end{tabular}
\floatfoot{
Note:  This table uses a linear adjustment of primary outcomes for program cost to compare HD and GD at exactly equivalent costs.  The \emph{Transfer value} column estimates the marginal effect of spending an extra \$100 through cash transfers.  The \emph{Cost-equivalent GD impact} column is estimated as a dummy for either HD or GD treatment, and estimates the impact of cash at the exact cost of HD.  The \emph{Differential impact of HD} column then estimates the differential effect of HD above cash at this benchmarked cost. Regressions include but do not report the lagged dependent variable, fixed effects for randomization blocks, and a set of LASSO-selected baseline covariates.  Standard errors are (in soft brackets) are clustered at the household level to reflect the design effect, and $p$-values corrected for False Discovery Rates across all the outcomes in the table are presented in hard brackets.  Stars on coefficient estimates are derived from the FDR-corrected $p$-values, *=10\%, **=5\%, and ***=1\% significance.  Employed is a dummy variable for spending more than 10 hours per week working for a wage or as primary operator of a microenterprise.  Productive hours are measured over prior 7 days in all activities other than own-farm agriculture.  Monthly income, productive assets, and household consumption are winsorized at 1\% and 99\% and analyzed in Inverse Hyperbolic Sine, meaning that treatment effects can be interpreted as percent changes. 
}
\end{footnotesize}
\end{table}

\cleardoublepage

\begin{table}[h]\centering
\caption{{\bf Benefit-Cost Ratios, Primary Outcomes}.}
     \begin{footnotesize}
 \label{t:benefit_cost_ratios_primary}
 \begin{tabular}{l *{6}{S[table-format=03.3]} SSSS}
\toprule
 & & \multicolumn{4}{c}{GiveDirectly} & & \multicolumn{4}{c}{$ p$-values} \\ 
\cmidrule(lr){3-6} \cmidrule(lr){8-11}
\multicolumn{1}{c}{\text{ }} & \multicolumn{1}{c}{\text{HD}} & \multicolumn{1}{c}{\text{Lower}} & \multicolumn{1}{c}{\text{Middle}} & \multicolumn{1}{c}{\text{Upper}} & \multicolumn{1}{c}{\text{Large}} & \multicolumn{1}{c}{\text{Combined}} & \multicolumn{1}{c}{\text{(a)}} & \multicolumn{1}{c}{\text{(b)}} & \multicolumn{1}{c}{\text{(c)}} & \multicolumn{1}{c}{\text{(d)}}\\
\midrule
\multirow[t]{ 2 }{0.15\textwidth}{Employed }  &     0.008 &     0.007 &    -0.004 &    -0.005 &     0.002 &     0.001 &      0.61 &      0.91 &      0.71 &      0.79 \\ 
 & (    0.010)  & (    0.012)  & (    0.010)  & (    0.007)  & (    0.005)  & (    0.005)  \\ \addlinespace[1ex] 
\multirow[t]{ 2 }{0.15\textwidth}{Productive hours }  &     0.999 &     0.088 &     0.129 &     0.095 &     0.274 &     0.249 &      0.36 &      0.13 &      0.74 &      0.93 \\ 
 & (    0.460)  & (    0.550)  & (    0.450)  & (    0.378)  & (    0.251)  & (    0.243)  \\ \addlinespace[1ex] 
\multirow[t]{ 2 }{0.15\textwidth}{Monthly income }  &     0.060 &     0.043 &     0.118 &    -0.039 &     0.084 &     0.025 &      0.35 &      0.86 &      0.65 &      0.21 \\ 
 & (    0.080)  & (    0.094)  & (    0.072)  & (    0.065)  & (    0.039)  & (    0.041)  \\
\bottomrule
\end{tabular}
\floatfoot{
    Note:  Table gives the impact per \$100 spent, which is calculated by dividing the estimated ITT impacts by the cost per arm in hundreds of dollars.  The standard errors in the table are similarly the ITT standard errors divided by costs.  Reported $p$-values in final three columns derived from $F$-tests of hypotheses that cost-benefit ratios are equal between: (a) joint test across all arms, (b) GD Lower and HD; (c) GD Lower and GD Large; and (d) GD Large and Combined arms.
}
\end{footnotesize}
\end{table}

\begin{landscape}

\begin{table}[h]\centering
\caption{{\bf Beneficiary Business Birth and Death}.}
     \begin{footnotesize}
 \label{t:business_survival}
 \begin{tabular}{l *{9}{c}}
\toprule
 & & \multicolumn{4}{c}{GiveDirectly} & & \multicolumn{1}{c}{\raisebox{-1ex}[0pt]{Control}} \\ 
\multicolumn{1}{c}{\text{ }} & \multicolumn{1}{c}{\text{HD}} & \multicolumn{1}{c}{\text{Lower}} & \multicolumn{1}{c}{\text{Middle}} & \multicolumn{1}{c}{\text{Upper}} & \multicolumn{1}{c}{\text{Large}} & \multicolumn{1}{c}{\text{Combined}} & \multicolumn{1}{c}{\text{Mean}} & \multicolumn{1}{c}{\text{Obs.}} & \multicolumn{1}{c}{\text{$ R^2$}}\\
\midrule
\multirow[t]{ 3 }{0.15\textwidth}{New business at midline }  &      0.22\ensuremath{^{***}} &      0.55\ensuremath{^{***}} &      0.55\ensuremath{^{***}} &      0.61\ensuremath{^{***}} &      0.64\ensuremath{^{***}} &      0.65\ensuremath{^{***}} &      0.50 &      1770 &      0.12 \\ 
 & (0.05)  & (0.08)  & (0.08)  & (0.08)  & (0.08)  & (0.07)  \\  & [0.00]  & [0.00]  & [0.00]  & [0.00]  & [0.00]  & [0.00]  \\ \addlinespace[1ex] 
\multirow[t]{ 3 }{0.15\textwidth}{New business at endline }  &      0.05 &      0.01 &      0.05 &      0.05 &      0.16\ensuremath{^{***}} &      0.10\ensuremath{^{**}} &      0.24 &      1822 &      0.02 \\ 
 & (0.03)  & (0.04)  & (0.04)  & (0.04)  & (0.05)  & (0.05)  \\  & [0.11]  & [0.53]  & [0.22]  & [0.19]  & [0.00]  & [0.05]  \\
\bottomrule
\end{tabular}

\floatfoot{
Notes:  Table uses the panel of firms reported on in the beneficiary survey, totalling variables within each round.  The first two rows count the number of new firms in each round that had not existed in the prior round, by beneficiary, and then impute zeros for individuals who reported no firms in the survey.  Rows 3-5 then take the universe of individuals who reported on any firm in the midline, and count the outcomes for those firms at endline; whether they were no longer operational (died), existed but were inoperative, or were operative.  Standard errors are (in soft brackets) are clustered at the household level to reflect the design effect, and $p$-values corrected for False Discovery Rates across all the outcomes in the table are presented in hard brackets.  Stars on coefficient estimates are derived from the FDR-corrected $p$-values, *=10\%, **=5\%, and ***=1\% significance.  
}
\end{footnotesize}
\end{table}

\cleardoublepage

\begin{table}[h]\centering
\caption{{\bf Midline Beneficiary enterprise analysis}.}
     \begin{footnotesize}
 \label{t:business_outcomes_enterprise_midline}
 \begin{tabular}{l *{9}{c}}
\toprule
 & & \multicolumn{4}{c}{GiveDirectly} & & \multicolumn{1}{c}{\raisebox{-1ex}[0pt]{Control}} \\ 
\multicolumn{1}{c}{\text{ }} & \multicolumn{1}{c}{\text{HD}} & \multicolumn{1}{c}{\text{Lower}} & \multicolumn{1}{c}{\text{Middle}} & \multicolumn{1}{c}{\text{Upper}} & \multicolumn{1}{c}{\text{Large}} & \multicolumn{1}{c}{\text{Combined}} & \multicolumn{1}{c}{\text{Mean}} & \multicolumn{1}{c}{\text{Obs.}} & \multicolumn{1}{c}{\text{$ R^2$}}\\
\midrule
\multirow[t]{ 3 }{0.15\textwidth}{Operating Businesses }  &      0.14\ensuremath{^{***}} &      0.44\ensuremath{^{***}} &      0.55\ensuremath{^{***}} &      0.62\ensuremath{^{***}} &      0.65\ensuremath{^{***}} &      0.63\ensuremath{^{***}} &      0.79 &      1770 &      0.10 \\ 
 & (0.06)  & (0.09)  & (0.10)  & (0.09)  & (0.10)  & (0.08)  \\  & [0.01]  & [0.00]  & [0.00]  & [0.00]  & [0.00]  & [0.00]  \\ \addlinespace[1ex] 
\multirow[t]{ 3 }{0.15\textwidth}{Owned Businesses }  &      0.14\ensuremath{^{***}} &      0.48\ensuremath{^{***}} &      0.57\ensuremath{^{***}} &      0.60\ensuremath{^{***}} &      0.66\ensuremath{^{***}} &      0.66\ensuremath{^{***}} &      0.71 &      1770 &      0.11 \\ 
 & (0.06)  & (0.08)  & (0.09)  & (0.09)  & (0.09)  & (0.07)  \\  & [0.01]  & [0.00]  & [0.00]  & [0.00]  & [0.00]  & [0.00]  \\
\bottomrule
\end{tabular}

\floatfoot{
Notes:  Table analyzes the results of the midline beneficiary enterprise survey, totalling variables across all businesses operated by a given beneficiary and then imputing zeros to survey respondents with no businesses.  Analysis is weighted using attrition weights. Standard errors are (in soft brackets) are clustered at the household level to reflect the design effect, and $p$-values corrected for False Discovery Rates across all the outcomes in the table are presented in hard brackets.  Stars on coefficient estimates are derived from the FDR-corrected $p$-values, *=10\%, **=5\%, and ***=1\% significance.  
}
\end{footnotesize}
\end{table}

\begin{table}[h]\centering
\caption{{\bf Endline Beneficiary enterprise analysis}.}
     \begin{footnotesize}
 \label{t:business_outcomes_enterprise_endline}
 \begin{tabular}{l *{9}{c}}
\toprule
 & & \multicolumn{4}{c}{GiveDirectly} & & \multicolumn{1}{c}{\raisebox{-1ex}[0pt]{Control}} \\ 
\multicolumn{1}{c}{\text{ }} & \multicolumn{1}{c}{\text{HD}} & \multicolumn{1}{c}{\text{Lower}} & \multicolumn{1}{c}{\text{Middle}} & \multicolumn{1}{c}{\text{Upper}} & \multicolumn{1}{c}{\text{Large}} & \multicolumn{1}{c}{\text{Combined}} & \multicolumn{1}{c}{\text{Mean}} & \multicolumn{1}{c}{\text{Obs.}} & \multicolumn{1}{c}{\text{$ R^2$}}\\
\midrule
\multirow[t]{ 3 }{0.15\textwidth}{Operating Businesses }  &      0.10\ensuremath{^{**}} &      0.18\ensuremath{^{**}} &      0.27\ensuremath{^{***}} &      0.29\ensuremath{^{***}} &      0.42\ensuremath{^{***}} &      0.33\ensuremath{^{***}} &      0.40 &      1822 &      0.05 \\ 
 & (0.04)  & (0.07)  & (0.07)  & (0.07)  & (0.08)  & (0.06)  \\  & [0.03]  & [0.01]  & [0.00]  & [0.00]  & [0.00]  & [0.00]  \\ \addlinespace[1ex] 
\multirow[t]{ 3 }{0.15\textwidth}{Owned Businesses }  &      0.11\ensuremath{^{**}} &      0.17\ensuremath{^{**}} &      0.27\ensuremath{^{***}} &      0.30\ensuremath{^{***}} &      0.44\ensuremath{^{***}} &      0.34\ensuremath{^{***}} &      0.43 &      1822 &      0.05 \\ 
 & (0.05)  & (0.07)  & (0.07)  & (0.07)  & (0.08)  & (0.07)  \\  & [0.02]  & [0.02]  & [0.00]  & [0.00]  & [0.00]  & [0.00]  \\
\bottomrule
\end{tabular}
\floatfoot{
Notes:  Table analyzes the results of the endline beneficiary enterprise survey, totalling variables across all businesses operated by a given beneficiary and then imputing zeros to survey respondents with no businesses.  Standard errors are (in soft brackets) are clustered at the household level to reflect the design effect, and $p$-values corrected for False Discovery Rates across all the outcomes in the table are presented in hard brackets.  Stars on coefficient estimates are derived from the FDR-corrected $p$-values, *=10\%, **=5\%, and ***=1\% significance.  
}
\end{footnotesize}
\end{table}

\cleardoublepage

\end{landscape}

\begin{table}[h]\centering
\caption{{\bf Aggregating cash flows over the study period}.}
     \begin{footnotesize}
 \label{t:cash_accounting}
 \begin{tabular}{l *{7}{S}}
\toprule
&  &\multicolumn{5}{c}{Cash transfers} \\ 
\cmidrule(lr){4-7} 
\multicolumn{1}{c}{\text{ }} & \multicolumn{1}{c}{\text{Control mean}} & \multicolumn{1}{c}{\text{HD}} & \multicolumn{1}{c}{\text{Lower}} & \multicolumn{1}{c}{\text{Middle}} & \multicolumn{1}{c}{\text{Upper}} & \multicolumn{1}{c}{\text{Large}} & \multicolumn{1}{c}{\text{Combined}}\\
\midrule
\multicolumn{8}{l}{\emph{Panel A.  Flow impacts through midline}}  \\ 
\addlinespace[1ex] Cash received  &      0.00 &      0.00 &    317.16 &    410.65 &    502.96 &    750.30 &    410.65 \\ 
\addlinespace[1ex] 
Beneficiary income  &    209.36 &     65.73 &    158.39 &    226.36 &    239.48 &    152.48 &    218.09 \\ 
\addlinespace[1ex] 
Transfers received  &     23.38 &     -7.78 &     40.91 &     62.42 &     61.20 &     72.71 &     57.55 \\ 
\addlinespace[1ex] 
Household consumption  &    625.85 &     34.19 &    124.50 &    166.52 &    146.09 &    223.07 &    169.13 \\ 
\addlinespace[1ex] 
Loans made  &      3.83 &      0.43 &      3.46 &      1.74 &      5.41 &      2.61 &      4.74 \\
\bottomrule
\end{tabular}

    \floatfoot{
        Notes.  Table presents control means and estimated impacts on financial values, in dollars.  Flow consumption is measured in the survey monthly, and here we use the midline treatment effect for the first 12 months and the endline effect for the subsequent 28 months.  Inter-household flows are measured with an annual recall, and we take a similar approach, using midline estimates for the midline period and endline estimates for the period between midline and endline.   All other variables are stocks measured at follow-up. \emph{Total income} is the sum of cash received, beneficiary income, and transfers received.  \emph{Total expenditure} is the sum of household consumption, loans made, and transfers made.  \emph{Total stock values} are the sum of  livestock values, other productive asset values, savings values, and the negative of debt values.  Randomization inference $p$-values, in brackets, from test of null of no cumulative effect of each arm on income. \emph{Share accounted} is the ratio of the sum of total outflows plus stock values to total inflows.  
    }
\end{footnotesize}
\end{table}

\clearpage

\cleardoublepage

\begin{figure}[!p]
\caption{{\bf Graphical Representation of Spillovers}}
\label{f:interference_combined}
\begin{center}
\includegraphics[width=0.95\linewidth]{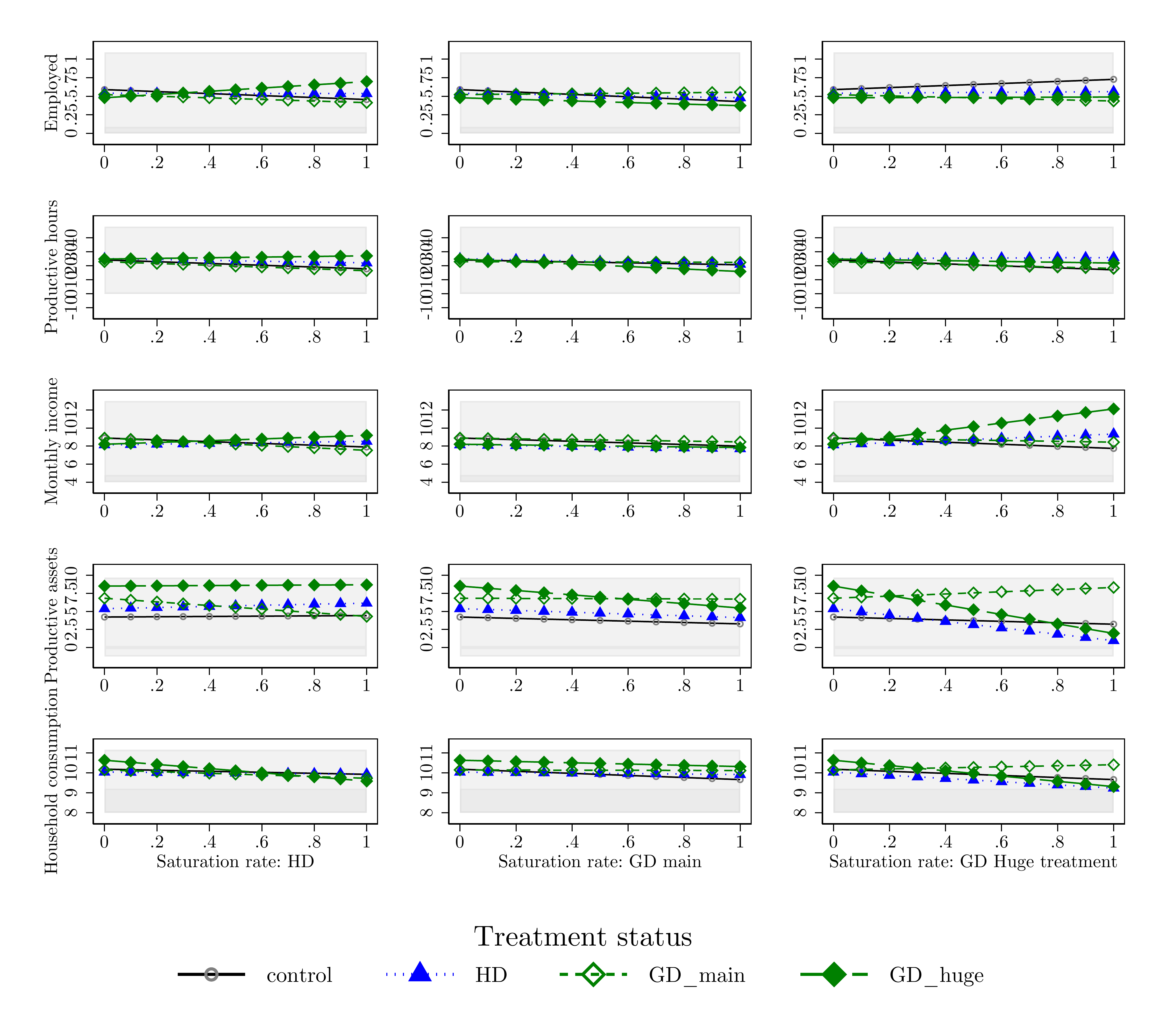}
\end{center}

\floatfoot{ 
\begin{footnotesize}
Notes:  Each panel presents predicted outcomes under each of the four main treatment arms (Control, HD, GD Main, and GD Large), as the saturation level of a specific active treatment arm changes.  Rows correspond to the outcomes of employment, productive hours, and the inverse hyperbolic sine of monthly incomes, productive assets, and household consumption per adult-equivalent, respectively. Horizontal shaded bands highlight one standard deviation above and below the control mean.  Columns illustrate effects of variation in saturation rates in HD, GD-main, and GD-large, respectively.  All predicted outcomes evaluated at means of covariates used in the estimating equation.
\end{footnotesize}
}
\end{figure}

\cleardoublepage

%--------------------------------------------------------------------------%
%       Appendices                                                         %
%--------------------------------------------------------------------------%
\cleardoublepage
\appendix
\noappendicestocpagenum \addappheadtotoc
\makeatletter
\def\@seccntformat#1{Appendix\ \csname the#1\endcsname\quad}
\def\@subseccntformat#1{\csname the#1\endcsname\quad}
\makeatother
\renewcommand\thetable{\Alph{section}.\arabic{table}}
\renewcommand\thefigure{\Alph{section}.\arabic{figure}}

\counterwithin{figure}{section}
\counterwithin{table}{section}

\cleardoublepage

%!TEX root=./ms.tex

\cleardoublepage
\begin{landscape}
\section{Appendix Tables}

    \vspace*{\fill}
\begin{table}[!h]\centering
\caption{{\bf Study Design}}\label{t:treatment_assignment}
%!TEX root=./../HD_PAP.tex

\begin{tabular}{L{1in} *{7}{C{1in}}}
\toprule %
		&			&				& \multicolumn{4}{c}{GiveDirectly} & Combined \tabularnewline
\cmidrule(lr){4-7} \cmidrule(lr){8-8} %
Sector 	&  Control 	& Huguka Dukore & 317.16 & 410.65 & 502.96 & 750.30 & HD + 410.65 \tabularnewline 
\midrule %
Kaduha	&	63	&	60	&	21	&	21	&	22	&	22	&	26	\tabularnewline 
Kibumbwe	&	32	&	37	&	10	&	10	&	12	&	13	&	13	\tabularnewline 
Kigabiro	&	14	&	12	&	4	&	5	&	4	&	5	&	5	\tabularnewline 
Kiyumba	&	17	&	17	&	6	&	6	&	6	&	6	&	8	\tabularnewline 
Mugano	&	51	&	51	&	18	&	18	&	18	&	18	&	22	\tabularnewline 
Muhazi	&	39	&	40	&	13	&	19	&	13	&	18	&	17	\tabularnewline 
Munyaga	&	34	&	34	&	10	&	10	&	10	&	12	&	14	\tabularnewline 
Munyiginya	&	25	&	25	&	8	&	8	&	8	&	10	&	10	\tabularnewline 
Musange	&	30	&	29	&	10	&	10	&	10	&	9	&	12	\tabularnewline 
Mushishiro	&	24	&	23	&	6	&	6	&	6	&	9	&	8	\tabularnewline 
Nyakariro	&	49	&	50	&	16	&	17	&	19	&	17	&	22	\tabularnewline 
Nyarusange	&	57	&	54	&	21	&	20	&	19	&	19	&	24	\tabularnewline 
Shyogwe	&	53	&	53	&	18	&	18	&	18	&	20	&	22	\tabularnewline 
\midrule %
Total	&	488	&	485	&	161	&	168	&	165	&	178	&	203	\tabularnewline 
\bottomrule 
\end{tabular}
\floatfoot{ 
\begin{footnotesize}
Note :  This table gives the number of study individuals assigned to each treatment arm in each of the 13 sectors within which lotteries were conducted.  The lotteries were blocked so that fixed fractions of individuals are assigned to each arm.
\end{footnotesize}
}
\end{table}
\vspace*{\fill}
\end{landscape}

\begin{table}[h]\centering
\caption{{\bf Attrition:  Tracking at Endline by Treatment}.}
   \label{t:Attrition_simple}
 {
\def\sym#1{\ifmmode^{#1}\else\(^{#1}\)\fi}
\begin{tabular}{l*{3}{c}}
\toprule
                    &\multicolumn{1}{c}{\shortstack{Any  \\ Treatment}}&\multicolumn{1}{c}{\shortstack{Basic \\ Treatment}}&\multicolumn{1}{c}{\shortstack{Granular \\ Treatment}}\\
\midrule
Any Treatment       &     0.00594         &                     &                     \\
                    &   (0.00678)         &                     &                     \\
\addlinespace
HD                  &                     &     0.00607         &     0.00607         \\
                    &                     &   (0.00790)         &   (0.00790)         \\
\addlinespace
Treatment arm: Gd   &                     &     0.00654         &                     \\
                    &                     &   (0.00738)         &                     \\
\addlinespace
Combined            &                     &     0.00366         &     0.00366         \\
\bottomrule
\end{tabular}
}

\floatfoot{
    \begin{footnotesize}
    Notes:  Table examines overall attrition rates from the endline survey by treatment arm.  Sample is the entire baseline survey, outcome variable is a dummy for being successfully tracked at endline.  Covariates are the treatment arm for each individual.  Standard errors clustered at the household level, *=10\%, **=5\%, and ***=1\% significance.
\end{footnotesize}
}
\end{table}

\begin{table}[h]\centering
\caption{{\bf Attrition:  Tracking at Endline on Covariates}.}
     \begin{footnotesize}
 \label{t:Attrition}
 \begin{tabular}{l *{1}{S} ScSS}
\toprule
\multicolumn{1}{c}{\text{ }} & \multicolumn{1}{c}{\text{R3 tracked}} & \multicolumn{1}{c}{\text{Control mean}} & \multicolumn{1}{c}{\text{Observations}} & \multicolumn{1}{c}{\text{$ R^2$}}\\
\midrule
\multirow[t]{ 3 }{0.2\textwidth}{Ubudehe category I }  &     -0.10 &      0.33 &      1797 &      0.00 \\ 
 & (0.10)  \\  & [    1.00]  \\ \addlinespace[1ex] 
\multirow[t]{ 3 }{0.2\textwidth}{Beneficiary female }  &      0.18 &      0.60 &      1848 &      0.00 \\ 
 & (0.10)  \\  & [    0.89]  \\ \addlinespace[1ex] 
\multirow[t]{ 3 }{0.2\textwidth}{Beneficiary age }  &      0.17 &     23.36 &      1848 &      0.00 \\
\bottomrule
\end{tabular}
\floatfoot{
Notes:   Table correlates attrition from the endline survey with baseline covariates.  Sample is the entire baseline survey, outcome variable is a dummy for being successfully tracked at endline.  Standard errors are (in soft brackets) are clustered at the household level to reflect the design effect, and $p$-values corrected for False Discovery Rates across all the outcomes in the table are presented in hard brackets.  Stars on coefficient estimates are derived from the FDR-corrected $p$-values, *=10\%, **=5\%, and ***=1\% significance.  
}
\end{footnotesize}
\end{table}

\begin{table}[h]\centering
\caption{{\bf Balance using Endline Sample}.}
     \begin{footnotesize}
 \label{t:Balance}
 \begin{tabular}{l *{6}{S} ScSS}
\toprule
 & & \multicolumn{4}{c}{GiveDirectly} & & \multicolumn{1}{c}{\raisebox{-1ex}[0pt]{Control}}  \\ 
\cmidrule(lr){3-6}
\multicolumn{1}{c}{\text{ }} & \multicolumn{1}{c}{\text{HD}} & \multicolumn{1}{c}{\text{Lower}} & \multicolumn{1}{c}{\text{Middle}} & \multicolumn{1}{c}{\text{Upper}} & \multicolumn{1}{c}{\text{Large}} & \multicolumn{1}{c}{\text{Combined}} & \multicolumn{1}{c}{\text{Mean}} & \multicolumn{1}{c}{\text{Obs.}} & \multicolumn{1}{c}{\text{$ R^2$}} & \multicolumn{1}{c}{\text{$ p$-value}}\\
\midrule
\multirow[t]{ 3 }{0.2\textwidth}{Ubudehe category I }  &      0.01 &     -0.00 &      0.04 &      0.00 &      0.00 &     -0.03 &      0.32 &      1771 &      0.07 &      0.90 \\ 
 & (0.03)  & (0.05)  & (0.05)  & (0.04)  & (0.04)  & (0.04)  \\  & [    1.00]  & [    1.00]  & [    1.00]  & [    1.00]  & [    1.00]  & [    1.00]  \\ \addlinespace[1ex] 
\multirow[t]{ 3 }{0.2\textwidth}{Beneficiary female }  &      0.02 &     -0.02 &      0.06 &     -0.01 &      0.01 &     -0.03 &      0.60 &      1822 &      0.04 &      0.66 \\ 
 & (0.03)  & (0.04)  & (0.04)  & (0.04)  & (0.04)  & (0.04)  \\  & [    1.00]  & [    1.00]  & [    1.00]  & [    1.00]  & [    1.00]  & [    1.00]  \\
\bottomrule
\end{tabular}
\floatfoot{
Notes:  Table examines balance of the experiment across treatment arms using baseline covariates and the attrited sample that is used for endline analysis.  Standard errors are (in soft brackets) are clustered at the household level to reflect the design effect, and $p$-values corrected for False Discovery Rates across all the outcomes in the table are presented in hard brackets.  Stars on coefficient estimates are derived from the FDR-corrected $p$-values, *=10\%, **=5\%, and ***=1\% significance.  
}
\end{footnotesize}
\end{table}

\cleardoublepage

\begin{landscape}

\begin{table}[h]\centering
\caption{{\bf Simple ITT, Secondary Outcomes: Beneficiary welfare}.}
     \begin{footnotesize}
 \label{t:itt_fullspec_secondary_welfare}     
 \begin{tabular}{l *{6}{S} ScSSSSS}
\toprule
 & & \multicolumn{4}{c}{GiveDirectly} & & \multicolumn{1}{c}{\raisebox{-1ex}[0pt]{Control}} & & & \multicolumn{3}{c}{$ p$-values} \\ 
\cmidrule(lr){3-6} \cmidrule(lr){11-13}
\multicolumn{1}{c}{\text{ }} & \multicolumn{1}{c}{\text{HD}} & \multicolumn{1}{c}{\text{Lower}} & \multicolumn{1}{c}{\text{Middle}} & \multicolumn{1}{c}{\text{Upper}} & \multicolumn{1}{c}{\text{Large}} & \multicolumn{1}{c}{\text{Combined}} & \multicolumn{1}{c}{\text{Mean}} & \multicolumn{1}{c}{\text{Obs.}} & \multicolumn{1}{c}{\text{$ R^2$}} & \multicolumn{1}{c}{\text{(a)}} & \multicolumn{1}{c}{\text{(b)}} & \multicolumn{1}{c}{\text{(c)}}\\
\midrule
\multicolumn{13}{l}{\emph{Panel A. Midline}}  \\ 
\addlinespace[1ex] \multirow[t]{ 3 }{0.15\textwidth}{Subjective well-being }  &      0.18\ensuremath{^{***}} &      0.41\ensuremath{^{***}} &      0.53\ensuremath{^{***}} &      0.49\ensuremath{^{***}} &      0.55\ensuremath{^{***}} &      0.41\ensuremath{^{***}} &     -0.01 &      1770 &      0.13 &      0.04 &      0.08 &      0.19 \\ 
 & (0.07)  & (0.09)  & (0.10)  & (0.09)  & (0.09)  & (0.09)  \\  & [    0.00]  & [    0.00]  & [    0.00]  & [    0.00]  & [    0.00]  & [    0.00]  \\ \addlinespace[1ex] 
\multirow[t]{ 3 }{0.15\textwidth}{Mental health }  &     -0.04 &     -0.05 &      0.07 &      0.02 &      0.11 &      0.12 &      0.01 &      1770 &      0.06 &      0.92 &      0.27 &      0.87 \\ 
 & (0.07)  & (0.09)  & (0.10)  & (0.09)  & (0.10)  & (0.09)  \\
\bottomrule
\end{tabular}
\floatfoot{
    Notes:   Regressions include but do not report the lagged dependent variable, fixed effects for randomization blocks, and a set of LASSO-selected baseline covariates.  Standard errors are (in soft brackets) are clustered at the household level to reflect the design effect, and $p$-values corrected for False Discovery Rates across all the outcomes in the table are presented in hard brackets.  Stars on coefficient estimates are derived from the FDR-corrected $p$-values, *=10\%, **=5\%, and ***=1\% significance.  Reported $p$-values in final three columns derived from $F$-tests of hypotheses that cost-benefit ratios are equal between: (a) GD Lower and HD; (b) GD Lower and GD Large; and (c) GD Large and Combined treatments.
}
\end{footnotesize}
\end{table}

\begin{table}[h]\centering
\caption{{\bf Simple ITT, Secondary Outcomes: Skills}.}
     \begin{footnotesize}
 \label{t:itt_fullspec_secondary_skills}     
 \begin{tabular}{l *{6}{S} ScSSSSS}
\toprule
 & & \multicolumn{4}{c}{GiveDirectly} & & \multicolumn{1}{c}{\raisebox{-1ex}[0pt]{Control}} & & & \multicolumn{3}{c}{$ p$-values} \\ 
\cmidrule(lr){3-6} \cmidrule(lr){11-13}
\multicolumn{1}{c}{\text{ }} & \multicolumn{1}{c}{\text{HD}} & \multicolumn{1}{c}{\text{Lower}} & \multicolumn{1}{c}{\text{Middle}} & \multicolumn{1}{c}{\text{Upper}} & \multicolumn{1}{c}{\text{Large}} & \multicolumn{1}{c}{\text{Combined}} & \multicolumn{1}{c}{\text{Mean}} & \multicolumn{1}{c}{\text{Obs.}} & \multicolumn{1}{c}{\text{$ R^2$}} & \multicolumn{1}{c}{\text{(a)}} & \multicolumn{1}{c}{\text{(b)}} & \multicolumn{1}{c}{\text{(c)}}\\
\midrule
\multicolumn{13}{l}{\emph{Panel A. Midline}}  \\ 
\addlinespace[1ex] \multirow[t]{ 3 }{0.15\textwidth}{Aspirations }  &     -0.00 &      0.10 &     -0.04 &      0.14 &      0.03 &      0.16 &     -0.01 &      1770 &      0.08 &      0.35 &      0.36 &      0.17 \\ 
 & (0.07)  & (0.09)  & (0.09)  & (0.09)  & (0.09)  & (0.08)  \\  & [    1.00]  & [    0.43]  & [    0.75]  & [    0.27]  & [    0.88]  & [    0.23]  \\ \addlinespace[1ex] 
\multirow[t]{ 3 }{0.15\textwidth}{Business knowledge }  &      0.65\ensuremath{^{***}} &      0.08 &      0.06 &      0.06 &     -0.00 &      0.61\ensuremath{^{***}} &      0.00 &      1770 &      0.22 &      0.00 &      0.38 &      0.00 \\ 
 & (0.07)  & (0.10)  & (0.09)  & (0.09)  & (0.09)  & (0.09)  \\
\bottomrule
\end{tabular}
\floatfoot{
    Notes:   Regressions include but do not report the lagged dependent variable, fixed effects for randomization blocks, and a set of LASSO-selected baseline covariates.  Standard errors are (in soft brackets) are clustered at the household level to reflect the design effect, and $p$-values corrected for False Discovery Rates across all the outcomes in the table are presented in hard brackets.  Stars on coefficient estimates are derived from the FDR-corrected $p$-values, *=10\%, **=5\%, and ***=1\% significance.  Reported $p$-values in final three columns derived from $F$-tests of hypotheses that cost-benefit ratios are equal between: (a) GD Lower and HD; (b) GD Lower and GD Large; and (c) GD Large and Combined treatments.
}
\end{footnotesize}
\end{table}

\begin{table}[h]\centering
\caption{{\bf Moving across villages}.}
     \begin{footnotesize}
 \label{t:res_status}
 \begin{tabular}{l *{9}{c}}
\toprule
 & & \multicolumn{4}{c}{GiveDirectly} & & \multicolumn{1}{c}{\raisebox{-1ex}[0pt]{Control}} & &  \\ 
\multicolumn{1}{c}{\text{ }} & \multicolumn{1}{c}{\text{HD}} & \multicolumn{1}{c}{\text{Lower}} & \multicolumn{1}{c}{\text{Middle}} & \multicolumn{1}{c}{\text{Upper}} & \multicolumn{1}{c}{\text{Large}} & \multicolumn{1}{c}{\text{Combined}} & \multicolumn{1}{c}{\text{Mean}} & \multicolumn{1}{c}{\text{Obs.}} & \multicolumn{1}{c}{\text{$ R^2$}}\\
\midrule
\multirow[t]{ 3 }{0.15\textwidth}{Urban R2 }  &     -0.00 &      0.04 &     -0.03 &     -0.04 &     -0.04 &     -0.03 &      0.18 &      1770 &      0.24 \\ 
 & (0.02)  & (0.04)  & (0.03)  & (0.03)  & (0.03)  & (0.03)  \\  & [1.00]  & [0.85]  & [0.88]  & [0.80]  & [0.80]  & [0.88]  \\ \addlinespace[1ex] 
\multirow[t]{ 3 }{0.15\textwidth}{New Village R2 }  &      0.05 &      0.06 &      0.07 &      0.02 &      0.07 &      0.05 &      0.16 &      1770 &      0.02 \\ 
 & (0.03)  & (0.04)  & (0.04)  & (0.04)  & (0.04)  & (0.04)  \\  & [0.74]  & [0.80]  & [0.80]  & [0.88]  & [0.74]  & [0.80]  \\ \addlinespace[1ex] 
\multirow[t]{ 3 }{0.15\textwidth}{Urban R3 }  &     -0.01 &      0.03 &     -0.02 &     -0.06 &     -0.07 &     -0.05 &      0.22 &      1822 &      0.16 \\ 
 & (0.03)  & (0.04)  & (0.04)  & (0.03)  & (0.03)  & (0.03)  \\  & [0.98]  & [0.88]  & [0.88]  & [0.74]  & [0.62]  & [0.80]  \\ \addlinespace[1ex] 
\multirow[t]{ 3 }{0.15\textwidth}{New Village R3 }  &      0.01 &      0.04 &      0.06 &      0.04 &      0.02 &     -0.01 &      0.37 &      1822 &      0.02 \\ 
 & (0.03)  & (0.05)  & (0.04)  & (0.04)  & (0.04)  & (0.04)  \\  & [0.88]  & [0.88]  & [0.80]  & [0.88]  & [0.88]  & [1.00]  \\ \addlinespace[1ex] 
\bottomrule
\end{tabular}
\floatfoot{
Notes:  Table analyzes the extent to which the beneficiary had moved across villages at midline (R2) or endline (R3).  `Urban' is a dummy variable indicating that the village in which the beneficiary resides in that round is classified as semi-urban, peri-urban, or urban (rather than rural).  `New Village' is a dummy for the village being a different one than the baseline village.  Standard errors are (in soft brackets) are clustered at the household level to reflect the design effect, and $p$-values corrected for False Discovery Rates across all the outcomes in the table are presented in hard brackets.  Stars on coefficient estimates are derived from the FDR-corrected $p$-values, *=10\%, **=5\%, and ***=1\% significance.  
}
\end{footnotesize}
\end{table}

\cleardoublepage

\begin{table}[h]\centering
\caption{{\bf Education and Time Use}.}
     \begin{footnotesize}
 \label{t:education_time}
 \begin{tabular}{l *{9}{c}}
\toprule
 & & \multicolumn{4}{c}{GiveDirectly} & & \multicolumn{1}{c}{\raisebox{-1ex}[0pt]{Control}} \\ 
\multicolumn{1}{c}{\text{ }} & \multicolumn{1}{c}{\text{HD}} & \multicolumn{1}{c}{\text{Lower}} & \multicolumn{1}{c}{\text{Middle}} & \multicolumn{1}{c}{\text{Upper}} & \multicolumn{1}{c}{\text{Large}} & \multicolumn{1}{c}{\text{Combined}} & \multicolumn{1}{c}{\text{Mean}} & \multicolumn{1}{c}{\text{Obs.}} & \multicolumn{1}{c}{\text{$ R^2$}}\\
\midrule
\multirow[t]{ 3 }{0.15\textwidth}{Highest Grade }  &      0.28 &     -0.14 &     -0.14 &      0.51 &      0.36 &      0.00 &     12.02 &      1822 &      0.05 \\ 
 & (0.25)  & (0.35)  & (0.33)  & (0.33)  & (0.34)  & (0.33)  \\  & [1.00]  & [1.00]  & [1.00]  & [1.00]  & [1.00]  & [1.00]  \\ \addlinespace[1ex] 
\multirow[t]{ 3 }{0.15\textwidth}{Hours in School }  &     -0.68 &      0.25 &     -0.88 &     -0.42 &     -0.52 &     -0.47 &      3.05 &      1822 &      0.01 \\ 
 & (0.45)  & (0.66)  & (0.51)  & (0.65)  & (0.58)  & (0.60)  \\  & [1.00]  & [1.00]  & [1.00]  & [1.00]  & [1.00]  & [1.00]  \\
\bottomrule
\end{tabular}
\floatfoot{
Notes:  Table analyzes endline education and time use variables.  Highest grade is an ordinal variable measuring completed schooling with the control mean representing one year of post-primary education.   `Hours in School' and `Hours Domestic Work' give the number of hours over the seven days prior to the endline that the respondent reports spending in each activity.  `Reservation wages' give the survey response to the daily wage the respendent said they would need to be paid to take a job in their village and in the nearest town, respectively (USD).  Standard errors are (in soft brackets) are clustered at the household level to reflect the design effect, and $p$-values corrected for False Discovery Rates across all the outcomes in the table are presented in hard brackets.  Stars on coefficient estimates are derived from the FDR-corrected $p$-values, *=10\%, **=5\%, and ***=1\% significance.  
}
\end{footnotesize}
\end{table}

\begin{table}[h]\centering
\caption{{\bf Midline Household Enterprise Analysis}.}
     \begin{footnotesize}
 \label{t:business_outcomes_hh_midline}
 \begin{tabular}{l *{9}{c}}
\toprule
 & & \multicolumn{4}{c}{GiveDirectly} & & \multicolumn{1}{c}{\raisebox{-1ex}[0pt]{Control}} \\ 
\multicolumn{1}{c}{\text{ }} & \multicolumn{1}{c}{\text{HD}} & \multicolumn{1}{c}{\text{Lower}} & \multicolumn{1}{c}{\text{Middle}} & \multicolumn{1}{c}{\text{Upper}} & \multicolumn{1}{c}{\text{Large}} & \multicolumn{1}{c}{\text{Combined}} & \multicolumn{1}{c}{\text{Mean}} & \multicolumn{1}{c}{\text{Obs.}} & \multicolumn{1}{c}{\text{$ R^2$}}\\
\midrule
\multirow[t]{ 3 }{0.15\textwidth}{Owned Businesses }  &     -0.02 &     -0.01 &      0.04 &      0.00 &     -0.02 &     -0.01 &      0.06 &      1770 &      0.02 \\ 
 & (0.02)  & (0.03)  & (0.04)  & (0.02)  & (0.02)  & (0.02)  \\  & [0.92]  & [1.00]  & [1.00]  & [1.00]  & [1.00]  & [1.00]  \\ \addlinespace[1ex] 
\multirow[t]{ 3 }{0.15\textwidth}{Household Employees }  &     -0.05 &     -0.08 &     -0.03 &     -0.03 &     -0.08 &     -0.08 &      0.12 &      1770 &      0.02 \\ 
 & (0.05)  & (0.04)  & (0.06)  & (0.05)  & (0.04)  & (0.04)  \\  & [1.00]  & [0.92]  & [1.00]  & [1.00]  & [0.92]  & [0.92]  \\
\bottomrule
\end{tabular}
\floatfoot{
Notes:  Table analyzes the results of the midline household enterprise survey, totalling variables across all businesses operated within the household (other than by the beneficiary) and then imputing zeros to households with no businesses.  Analysis is weighted using attrition weights. Standard errors are (in soft brackets) are clustered at the household level to reflect the design effect, and $p$-values corrected for False Discovery Rates across all the outcomes in the table are presented in hard brackets.  Stars on coefficient estimates are derived from the FDR-corrected $p$-values, *=10\%, **=5\%, and ***=1\% significance.  
}
\end{footnotesize}
\end{table}

\begin{table}[h]\centering
\caption{{\bf Endline Household Enterprise Analysis}.}
     \begin{footnotesize}
 \label{t:business_outcomes_hh_endline}
 \begin{tabular}{l *{9}{c}}
\toprule
 & & \multicolumn{4}{c}{GiveDirectly} & & \multicolumn{1}{c}{\raisebox{-1ex}[0pt]{Control}} \\ 
\multicolumn{1}{c}{\text{ }} & \multicolumn{1}{c}{\text{HD}} & \multicolumn{1}{c}{\text{Lower}} & \multicolumn{1}{c}{\text{Middle}} & \multicolumn{1}{c}{\text{Upper}} & \multicolumn{1}{c}{\text{Large}} & \multicolumn{1}{c}{\text{Combined}} & \multicolumn{1}{c}{\text{Mean}} & \multicolumn{1}{c}{\text{Obs.}} & \multicolumn{1}{c}{\text{$ R^2$}}\\
\midrule
\multirow[t]{ 3 }{0.15\textwidth}{Owned Businesses }  &      0.01 &      0.01 &      0.02 &     -0.01 &      0.03 &     -0.00 &      0.06 &      1822 &      0.01 \\ 
 & (0.02)  & (0.02)  & (0.02)  & (0.02)  & (0.02)  & (0.02)  \\  & [1.00]  & [1.00]  & [1.00]  & [1.00]  & [0.90]  & [1.00]  \\ \addlinespace[1ex] 
\multirow[t]{ 3 }{0.15\textwidth}{Household Employees }  &     -0.02 &     -0.08 &     -0.02 &     -0.05 &     -0.02 &     -0.07 &      0.10 &      1822 &      0.02 \\ 
 & (0.03)  & (0.03)  & (0.04)  & (0.03)  & (0.05)  & (0.03)  \\  & [1.00]  & [0.26]  & [1.00]  & [0.56]  & [1.00]  & [0.36]  \\
\bottomrule
\end{tabular}
\floatfoot{
Notes:  Table analyzes the results of the endline household enterprise survey, totalling variables across all businesses operated within the household (other than by the beneficiary) and then imputing zeros to households with no businesses.  Standard errors are (in soft brackets) are clustered at the household level to reflect the design effect, and $p$-values corrected for False Discovery Rates across all the outcomes in the table are presented in hard brackets.  Stars on coefficient estimates are derived from the FDR-corrected $p$-values, *=10\%, **=5\%, and ***=1\% significance.  
}
\end{footnotesize}
\end{table}

\cleardoublepage

\end{landscape}

\begin{table}[h]\centering
\caption{{\bf Endline Beneficiary Enterprise with Gender Interactions}.}
     \begin{footnotesize}
 \label{t:beneficiary_endline_gender_int}
 \begin{tabular}{l *{8}{S}}
\toprule
\multicolumn{1}{c}{\text{ }} & \multicolumn{1}{c}{\text{\makecell[b]{Operating\\Business}}} & \multicolumn{1}{c}{\text{\makecell[b]{Own\\Business}}} & \multicolumn{1}{c}{\text{\makecell[b]{Household\\Employees}}} & \multicolumn{1}{c}{\text{\makecell[b]{NonHH\\Employees}}} & \multicolumn{1}{c}{\text{\makecell[b]{Days\\Worked}}} & \multicolumn{1}{c}{\text{\makecell[b]{Monthly\\Customers}}} & \multicolumn{1}{c}{\text{\makecell[b]{Daily\\Sales}}} & \multicolumn{1}{c}{\text{\makecell[b]{Monthly\\Profits}}}\\
\midrule
\multirow[t]{ 2}{0.2\textwidth}{HD} &      0.16 &      0.18 &      0.08 &      0.05 &      2.40 &      0.59 &      5.26 &      3.54 \\ 
 & (0.08)  & (0.08)  & (0.07)  & (0.08)  & (1.45)  & (    15.51)  & (2.15)  & (1.94)  \\ 
 & [0.18]  & [0.14]  & [0.58]  & [0.84]  & [0.36]  & [1.00]  & [0.10]  & [0.29]  \\ 
\addlinespace[1ex] \multirow[t]{ 2}{0.2\textwidth}{GD Lower} &      0.21 &      0.18 &      0.09 &      0.02 &      3.17 &     19.50 &      6.82 &      3.39 \\ 
 & (0.11)  & (0.12)  & (0.10)  & (0.09)  & (2.14)  & (    25.50)  & (3.65)  & (2.66)  \\ 
 & [0.29]  & [0.39]  & [0.63]  & [1.00]  & [0.40]  & [0.74]  & [0.29]  & [0.48]  \\ 
\addlinespace[1ex] \multirow[t]{ 2}{0.2\textwidth}{GD Mid} &      0.28\ensuremath{^{*}} &      0.27 &      0.02 &      0.17 &      3.35 &     -2.00 &      3.17 &      4.71 \\
\bottomrule
\end{tabular}
\floatfoot{
Notes: Table presents tests for heterogeneity of beneficiary enterprise effects by Gender.  Interacted coefficients  give the differential effect of each arm for women, `Female' gives the difference between women and men in the control group, and the uninteracted treatment terms give the impact of each arm for men.   Standard errors are (in soft brackets) are clustered at the household level to reflect the design effect, and $p$-values corrected for False Discovery Rates across all the outcomes in the table are presented in hard brackets.  Stars on coefficient estimates are derived from the FDR-corrected $p$-values, *=10\%, **=5\%, and ***=1\% significance.  
}
\end{footnotesize}
\end{table}

\cleardoublepage

\begin{table}[h]\centering
\caption{{\bf Standard Complementarities Test, Primary Outcomes}.}
     \begin{footnotesize}
 \label{t:complementarities_primary}
 \begin{tabular}{l *{3}{S} ScS}
\toprule
\multicolumn{1}{c}{\text{ }} & \multicolumn{1}{c}{\text{HD}} & \multicolumn{1}{c}{\text{GD}} & \multicolumn{1}{c}{\text{Complementarity}} & \multicolumn{1}{c}{\text{\makecell[b]{Control\\Mean}}} & \multicolumn{1}{c}{\text{Obs.}} & \multicolumn{1}{c}{\text{$ R^2$}}\\
\midrule
\multirow[t]{ 3 }{0.2\textwidth}{Employed }  &      0.04 &     -0.01 &     -0.02 &      0.50 &      1323 &      0.17 \\ 
 & (0.03)  & (0.05)  & (0.06)  \\  & [    0.67]  & [    1.00]  & [    1.00]  \\ \addlinespace[1ex] 
\multirow[t]{ 3 }{0.2\textwidth}{Productive hours }  &      3.78\ensuremath{^{*}} &      0.22 &     -1.80 &     19.43 &      1323 &      0.20 \\ 
 & (1.51)  & (2.21)  & (3.05)  \\  & [    0.06]  & [    1.00]  & [    1.00]  \\ \addlinespace[1ex] 
\multirow[t]{ 3 }{0.2\textwidth}{Monthly income }  &      0.27 &      0.54 &     -0.58 &      8.11 &      1323 &      0.16 \\ 
 & (0.27)  & (0.36)  & (0.52)  \\  & [    0.67]  & [    0.42]  & [    0.67]  \\ \addlinespace[1ex] 
\multirow[t]{ 3 }{0.2\textwidth}{Productive assets }  &      0.95\ensuremath{^{*}} &      2.35\ensuremath{^{***}} &     -0.26 &      3.90 &      1323 &      0.16 \\ 
 & (0.38)  & (0.56)  & (0.77)  \\  & [    0.06]  & [    0.00]  & [    1.00]  \\ \addlinespace[1ex] 
\multirow[t]{ 3 }{0.2\textwidth}{HH consumption per capita }  &      0.04 &      0.21 &     -0.08 &      9.84 &      1313 &      0.24 \\ 
 & (0.06)  & (0.10)  & (0.13)  \\  & [    1.00]  & [    0.14]  & [    1.00]  \\ \addlinespace[1ex] 
\bottomrule
\end{tabular}
\floatfoot{
  Notes:  Table tests for complementarities of cash and training on Primary outcomes.  `HD' is a modified dummy for receiving HD, and `GD' cash, including those in the Combined arm as having received both interventions.  The Complementarity dummy, for the Combined arm only, therefore tests whether the Combined arm produces an outcome different from what we would expect based on the sum of the HD and GD impacts alone.  Standard errors are (in soft brackets) are clustered at the household level to reflect the design effect, and $p$-values corrected for False Discovery Rates across all the outcomes in the table are presented in hard brackets.  Stars on coefficient estimates are derived from the FDR-corrected $p$-values, *=10\%, **=5\%, and ***=1\% significance. 
}
\end{footnotesize}
\end{table}

\begin{table}[h]\centering
\caption{{\bf Standard Complementarities Test, Secondary Outcomes}.}
     \begin{footnotesize}
 \label{t:complementarities_secondary}
 \begin{tabular}{l *{3}{S} ScS}
\toprule
\multicolumn{1}{c}{\text{ }} & \multicolumn{1}{c}{\text{HD}} & \multicolumn{1}{c}{\text{GD}} & \multicolumn{1}{c}{\text{Complementarity}} & \multicolumn{1}{c}{\text{\makecell[b]{Control\\Mean}}} & \multicolumn{1}{c}{\text{Obs.}} & \multicolumn{1}{c}{\text{$ R^2$}}\\
\midrule
\multicolumn{7}{l}{\emph{Panel A. Beneficiary welfare}}  \\ 
\addlinespace[1ex] \multirow[t]{ 3 }{0.2\textwidth}{Subjective well-being }  &      0.13 &      0.29\ensuremath{^{*}} &     -0.12 &      0.00 &      1323 &      0.11 \\ 
 & (0.07)  & (0.11)  & (0.14)  \\  & [    0.38]  & [    0.06]  & [    0.48]  \\ \addlinespace[1ex] 
\multirow[t]{ 3 }{0.2\textwidth}{Mental health }  &      0.06 &      0.11 &     -0.20 &     -0.00 &      1323 &      0.08 \\ 
 & (0.07)  & (0.09)  & (0.13)  \\  & [    0.48]  & [    0.46]  & [    0.45]  \\
\bottomrule
\end{tabular}
\floatfoot{
  Notes:  Table tests for complementarities of cash and training on Secondary outcomes.  `HD' is a modified dummy for receiving HD, and `GD' cash, including those in the Combined arm as having received both interventions.  The Complementarity dummy, for the Combined arm only, therefore tests whether the Combined arm produces an outcome different from what we would expect based on the sum of the HD and GD impacts alone.  Standard errors are (in soft brackets) are clustered at the household level to reflect the design effect, and $p$-values corrected for False Discovery Rates across all the outcomes in the table are presented in hard brackets.  Stars on coefficient estimates are derived from the FDR-corrected $p$-values, *=10\%, **=5\%, and ***=1\% significance. 
}
\end{footnotesize}
\end{table}

\begin{table}[h]\centering
\caption{{\bf Complementarities Test, Midline Business Outcomes}.}
     \begin{footnotesize}
 \label{t:complementarities_business_midline}
 \begin{tabular}{l *{3}{S} ScS}
\toprule
\multicolumn{1}{c}{\text{ }} & \multicolumn{1}{c}{\text{HD}} & \multicolumn{1}{c}{\text{GD}} & \multicolumn{1}{c}{\text{Complementarity}} & \multicolumn{1}{c}{\text{\makecell[b]{Control\\Mean}}} & \multicolumn{1}{c}{\text{Obs.}} & \multicolumn{1}{c}{\text{$ R^2$}}\\
\midrule
\multirow[t]{ 3 }{0.2\textwidth}{Operating Businesses }  &      0.12\ensuremath{^{*}} &      0.57\ensuremath{^{***}} &     -0.08 &      0.80 &      1289 &      0.09 \\ 
 & (0.06)  & (0.10)  & (0.13)  \\  & [    0.05]  & [    0.00]  & [    0.29]  \\ \addlinespace[1ex] 
\multirow[t]{ 3 }{0.2\textwidth}{Owned Businesses }  &      0.12\ensuremath{^{**}} &      0.59\ensuremath{^{***}} &     -0.08 &      0.72 &      1289 &      0.10 \\ 
 & (0.06)  & (0.09)  & (0.12)  \\  & [    0.03]  & [    0.00]  & [    0.29]  \\ \addlinespace[1ex] 
\multirow[t]{ 3 }{0.2\textwidth}{Household Employees }  &      0.16\ensuremath{^{**}} &      0.24\ensuremath{^{**}} &     -0.16 &      0.27 &      1289 &      0.03 \\
\bottomrule
\end{tabular}
\floatfoot{
  Notes:  Table tests for complementarities of cash and training on midline business outcomes.  `HD' is a modified dummy for receiving HD, and `GD' cash, including those in the Combined arm as having received both interventions.  The Complementarity dummy, for the Combined arm only, therefore tests whether the Combined arm produces an outcome different from what we would expect based on the sum of the HD and GD impacts alone.  Standard errors are (in soft brackets) are clustered at the household level to reflect the design effect, and $p$-values corrected for False Discovery Rates across all the outcomes in the table are presented in hard brackets.  Stars on coefficient estimates are derived from the FDR-corrected $p$-values, *=10\%, **=5\%, and ***=1\% significance. 
}
\end{footnotesize}
\end{table}

\begin{table}[h]\centering
\caption{{\bf Complementarities Test, Endline Business Outcomes}.}
     \begin{footnotesize}
 \label{t:complementarities_business_endline}
 \begin{tabular}{l *{3}{S} ScS}
\toprule
\multicolumn{1}{c}{\text{ }} & \multicolumn{1}{c}{\text{HD}} & \multicolumn{1}{c}{\text{GD}} & \multicolumn{1}{c}{\text{Complementarity}} & \multicolumn{1}{c}{\text{\makecell[b]{Control\\Mean}}} & \multicolumn{1}{c}{\text{Obs.}} & \multicolumn{1}{c}{\text{$ R^2$}}\\
\midrule
\multirow[t]{ 3 }{0.2\textwidth}{Operating Businesses }  &      0.10 &      0.27\ensuremath{^{***}} &     -0.03 &      0.40 &      1323 &      0.04 \\ 
 & (0.04)  & (0.07)  & (0.09)  \\  & [    0.12]  & [    0.00]  & [    0.73]  \\ \addlinespace[1ex] 
\multirow[t]{ 3 }{0.2\textwidth}{Owned Businesses }  &      0.11\ensuremath{^{*}} &      0.27\ensuremath{^{***}} &     -0.03 &      0.43 &      1323 &      0.04 \\ 
 & (0.05)  & (0.07)  & (0.10)  \\  & [    0.08]  & [    0.00]  & [    0.73]  \\ \addlinespace[1ex] 
\multirow[t]{ 3 }{0.2\textwidth}{Household Employees }  &      0.06 &      0.10 &     -0.02 &      0.17 &      1323 &      0.01 \\
\bottomrule
\end{tabular}
\floatfoot{
  Notes:  Table tests for complementarities of cash and training on endline business outcomes.  `HD' is a modified dummy for receiving HD, and `GD' cash, including those in the Combined arm as having received both interventions.  The Complementarity dummy, for the Combined arm only, therefore tests whether the Combined arm produces an outcome different from what we would expect based on the sum of the HD and GD impacts alone.  Standard errors are (in soft brackets) are clustered at the household level to reflect the design effect, and $p$-values corrected for False Discovery Rates across all the outcomes in the table are presented in hard brackets.  Stars on coefficient estimates are derived from the FDR-corrected $p$-values, *=10\%, **=5\%, and ***=1\% significance. 
}
\end{footnotesize}
\end{table}

\begin{table}[h]\centering
\caption{{\bf Cost Equivalent Analysis, Secondary Outcomes}.}
     \begin{footnotesize}
 \label{t:costequiv_secondary}
 \begin{tabular}{l *{3}{S} ScS}
\toprule
\multicolumn{1}{c}{\text{ }} & \multicolumn{1}{c}{\text{\makecell[b]{Differential impact \\ of HD}}} & \multicolumn{1}{c}{\text{\makecell[b]{Cost-equivalent\\GD impact}}} & \multicolumn{1}{c}{\text{\makecell[b]{Transfer\\Value}}} & \multicolumn{1}{c}{\text{\makecell[b]{Control\\Mean}}} & \multicolumn{1}{c}{\text{Obs.}} & \multicolumn{1}{c}{\text{$ R^2$}}\\
\midrule
\multicolumn{7}{l}{\emph{Panel A. Beneficiary welfare}}  \\ 
\addlinespace[1ex] \multirow[t]{ 3 }{0.2\textwidth}{Subjective well-being }  &     -0.15 &      0.29\ensuremath{^{***}} &     -0.01 &      0.00 &      1622 &      0.11 \\ 
 & (0.09)  & (0.08)  & (0.02)  \\  & [    0.22]  & [    0.01]  & [    0.76]  \\ \addlinespace[1ex] 
\multirow[t]{ 3 }{0.2\textwidth}{Mental health }  &     -0.02 &      0.08 &      0.01 &     -0.00 &      1622 &      0.07 \\ 
 & (0.08)  & (0.08)  & (0.02)  \\  & [    0.82]  & [    0.52]  & [    0.82]  \\
\bottomrule
\end{tabular}
\floatfoot{
Note:  This table uses a linear adjustment of secondary outcomes for program cost to compare HD and GD at exactly equivalent costs.  The \emph{Transfer value} column estimates the marginal effect of spending an extra \$100 through cash transfers.  The \emph{Cost-equivalent GD impact} column is estimated as a dummy for either HD or GD treatment, and estimates the impact of cash at the exact cost of HD.  The \emph{Differential impact of HD} column then estimates the differential effect of HD above cash at this benchmarked cost.  Regressions include but do not report the lagged dependent variable, fixed effects for randomization blocks, and a set of LASSO-selected baseline covariates.  Standard errors are (in soft brackets) are clustered at the household level to reflect the design effect, and $p$-values corrected for False Discovery Rates across all outcomes within each family are presented in hard brackets.  Stars on coefficient estimates are derived from the FDR-corrected $p$-values, *=10\%, **=5\%, and ***=1\% significance.  
}
\end{footnotesize}
\end{table}

\begin{landscape}

\begin{table}[h]\centering
\caption{{\bf Midline Business Cost Equivalence}.}
     \begin{footnotesize}
 \label{t:costequiv_business_midline}
 \begin{tabular}{l *{3}{S} ScS}
\toprule
\multicolumn{1}{c}{\text{ }} & \multicolumn{1}{c}{\text{\makecell[b]{Differential impact \\ of HD}}} & \multicolumn{1}{c}{\text{\makecell[b]{Cost-equivalent\\GD impact}}} & \multicolumn{1}{c}{\text{\makecell[b]{Transfer\\Value}}} & \multicolumn{1}{c}{\text{\makecell[b]{Control\\Mean}}} & \multicolumn{1}{c}{\text{Obs.}} & \multicolumn{1}{c}{\text{$ R^2$}}\\
\midrule
\multirow[t]{ 3 }{0.2\textwidth}{Operating Businesses }  &     -0.34\ensuremath{^{***}} &      0.46\ensuremath{^{***}} &      0.04\ensuremath{^{*}} &      0.80 &      1578 &      0.09 \\ 
 & (0.09)  & (0.09)  & (0.03)  \\  & [    0.00]  & [    0.00]  & [    0.06]  \\ \addlinespace[1ex] 
\multirow[t]{ 3 }{0.2\textwidth}{Owned Businesses }  &     -0.36\ensuremath{^{***}} &      0.49\ensuremath{^{***}} &      0.04\ensuremath{^{*}} &      0.72 &      1578 &      0.10 \\ 
 & (0.08)  & (0.08)  & (0.02)  \\  & [    0.00]  & [    0.00]  & [    0.06]  \\ \addlinespace[1ex] 
\multirow[t]{ 3 }{0.2\textwidth}{Household Employees }  &     -0.05 &      0.21\ensuremath{^{**}} &      0.05\ensuremath{^{*}} &      0.27 &      1578 &      0.04 \\
\bottomrule
\end{tabular}
\floatfoot{
Notes:  This table uses a linear adjustment of midline business outcomes for program cost to compare HD and GD at exactly equivalent costs.  The \emph{Transfer value} column estimates the marginal effect of spending an extra \$100 through cash transfers.  The \emph{Cost-equivalent GD impact} column is estimated as a dummy for either HD or GD treatment, and estimates the impact of cash at the exact cost of HD.  The \emph{Differential impact of HD} column then estimates the differential effect of HD above cash at this benchmarked cost.  Regressions include but do not report the lagged dependent variable, fixed effects for randomization blocks, and a set of LASSO-selected baseline covariates.  Standard errors are (in soft brackets) are clustered at the household level to reflect the design effect, and $p$-values corrected for False Discovery Rates across all outcomes within each family are presented in hard brackets.  Stars on coefficient estimates are derived from the FDR-corrected $p$-values, *=10\%, **=5\%, and ***=1\% significance. 
}
\end{footnotesize}
\end{table}

\begin{table}[h]\centering
\caption{{\bf Endline Business Cost Equivalence}.}
     \begin{footnotesize}
 \label{t:costequiv_business}
 \begin{tabular}{l *{3}{S} ScS}
\toprule
\multicolumn{1}{c}{\text{ }} & \multicolumn{1}{c}{\text{\makecell[b]{Differential impact \\ of HD}}} & \multicolumn{1}{c}{\text{\makecell[b]{Cost-equivalent\\GD impact}}} & \multicolumn{1}{c}{\text{\makecell[b]{Transfer\\Value}}} & \multicolumn{1}{c}{\text{\makecell[b]{Control\\Mean}}} & \multicolumn{1}{c}{\text{Obs.}} & \multicolumn{1}{c}{\text{$ R^2$}}\\
\midrule
\multirow[t]{ 3 }{0.2\textwidth}{Operating Businesses }  &     -0.07 &      0.17\ensuremath{^{*}} &      0.05\ensuremath{^{*}} &      0.40 &      1622 &      0.05 \\ 
 & (0.06)  & (0.06)  & (0.02)  \\  & [    0.32]  & [    0.08]  & [    0.08]  \\ \addlinespace[1ex] 
\multirow[t]{ 3 }{0.2\textwidth}{Owned Businesses }  &     -0.04 &      0.15\ensuremath{^{*}} &      0.06\ensuremath{^{*}} &      0.43 &      1622 &      0.05 \\ 
 & (0.07)  & (0.06)  & (0.02)  \\  & [    0.46]  & [    0.08]  & [    0.08]  \\ \addlinespace[1ex] 
\multirow[t]{ 3 }{0.2\textwidth}{Household Employees }  &     -0.04 &      0.09 &      0.02 &      0.17 &      1622 &      0.02 \\
\bottomrule
\end{tabular}
\floatfoot{
Notes:  This table uses a linear adjustment of endline business outcomes for program cost to compare HD and GD at exactly equivalent costs.  The \emph{Transfer value} column estimates the marginal effect of spending an extra \$100 through cash transfers.  The \emph{Cost-equivalent GD impact} column is estimated as a dummy for either HD or GD treatment, and estimates the impact of cash at the exact cost of HD.  The \emph{Differential impact of HD} column then estimates the differential effect of HD above cash at this benchmarked cost.  Regressions include but do not report the lagged dependent variable, fixed effects for randomization blocks, and a set of LASSO-selected baseline covariates.  Standard errors are (in soft brackets) are clustered at the household level to reflect the design effect, and $p$-values corrected for False Discovery Rates across all outcomes within each family are presented in hard brackets.  Stars on coefficient estimates are derived from the FDR-corrected $p$-values, *=10\%, **=5\%, and ***=1\% significance. 
}
\end{footnotesize}
\end{table}

\begin{table}[h]\centering
\caption{{\bf Robustness of Cost Equivalence Adjustment, Primary Outcomes}.}
     \begin{footnotesize}
\label{t:costequiv_linearity_primary}
 \begin{tabular}{l *{7}{S} ScS}
\toprule
\multicolumn{1}{c}{\text{ }} & \multicolumn{1}{c}{\text{\makecell[b]{Base\\Linear}}} & \multicolumn{1}{c}{\text{\makecell[b]{Quad-\\ratic}}} & \multicolumn{1}{c}{\text{\makecell[b]{Cubic\\}}} & \multicolumn{1}{c}{\text{\makecell[b]{Drop\\lower}}} & \multicolumn{1}{c}{\text{\makecell[b]{Drop\\mid}}} & \multicolumn{1}{c}{\text{\makecell[b]{Drop\\upper}}} & \multicolumn{1}{c}{\text{\makecell[b]{Drop\\huge}}}\\
\midrule
\multirow[t]{ 2 }{0.2\textwidth}{Employed }  &     0.030 &    -0.039 &    -0.063 &     0.076 &     0.020 &     0.018 &    -0.012 \\ 
 & (    0.042)  & (    0.068)  & (    0.146)  & (    0.057)  & (    0.047)  & (    0.044)  & (    0.056)  \\ \addlinespace[1ex] 
\multirow[t]{ 2 }{0.2\textwidth}{Productive hours }  &     3.412\ensuremath{^{*}} &     2.756 &     3.661 &     3.889 &     3.466 &     3.248 &     2.982 \\ 
 & (    2.069)  & (    3.401)  & (    7.041)  & (    2.867)  & (    2.293)  & (    2.160)  & (    2.764)  \\ \addlinespace[1ex] 
\multirow[t]{ 2 }{0.2\textwidth}{Monthly income }  &     0.111 &    -0.382 &     1.397 &     0.230 &     0.357 &    -0.089 &    -0.328 \\ 
 & (    0.332)  & (    0.553)  & (    1.146)  & (    0.447)  & (    0.384)  & (    0.343)  & (    0.452)  \\ \addlinespace[1ex] 
\multirow[t]{ 2 }{0.2\textwidth}{Productive assets }  &    -0.721 &     0.048 &     0.857 &    -1.287\ensuremath{^{*}} &    -0.511 &    -0.644 &    -0.277 \\ 
 & (    0.498)  & (    0.810)  & (    1.742)  & (    0.686)  & (    0.562)  & (    0.519)  & (    0.666)  \\ \addlinespace[1ex] 
\multirow[t]{ 2 }{0.2\textwidth}{HH consumption per capita }  &    -0.110 &    -0.058 &     0.077 &    -0.171 &    -0.081 &    -0.117 &    -0.070 \\ 
 & (    0.086)  & (    0.144)  & (    0.315)  & (    0.120)  & (    0.096)  & (    0.091)  & (    0.118)  \\ \addlinespace[1ex] 
\bottomrule
\end{tabular}
\floatfoot{
	Notes:  Table reports the coefficient on the differential effect of HD over cost-equivalent cash using seven different specifications.  Column 1 is the linear adjustment reported elsewhere.  Column 2 includes a quadratic, and column 3 a quadratic and cubic term in the cost deviations from Gikuriro.  Columns 4-7 leave out one of the cash treatment arms and repeat the linear cost adjustment.  Asterices denote significance at the 10, 5, and 1 percent levels, and are based on household-clustered standard errors, in parentheses. 
}
\end{footnotesize}
\end{table}

\begin{table}[h]\centering
\caption{{\bf Robustness of Cost Equivalence Adjustment, Secondary Outcomes}.}
     \begin{footnotesize}
 \label{t:costequiv_linearity_secondary}
 \begin{tabular}{l *{7}{S} ScS}
\toprule
\multicolumn{1}{c}{\text{ }} & \multicolumn{1}{c}{\text{\makecell[b]{Base\\Linear}}} & \multicolumn{1}{c}{\text{\makecell[b]{Quad-\\ratic}}} & \multicolumn{1}{c}{\text{\makecell[b]{Cubic\\}}} & \multicolumn{1}{c}{\text{\makecell[b]{Drop\\lower}}} & \multicolumn{1}{c}{\text{\makecell[b]{Drop\\mid}}} & \multicolumn{1}{c}{\text{\makecell[b]{Drop\\upper}}} & \multicolumn{1}{c}{\text{\makecell[b]{Drop\\huge}}}\\
\midrule
\multicolumn{8}{l}{\emph{Panel A. Beneficiary welfare}}  \\ 
\addlinespace[1ex] \multirow[t]{ 2 }{0.2\textwidth}{Subjective well-being }  &    -0.155\ensuremath{^{*}} &     0.098 &    -0.016 &    -0.293\ensuremath{^{**}} &    -0.160\ensuremath{^{*}} &    -0.105 &     0.013 \\ 
 & (    0.087)  & (    0.136)  & (    0.310)  & (    0.123)  & (    0.096)  & (    0.092)  & (    0.114)  \\ \addlinespace[1ex] 
\multirow[t]{ 2 }{0.2\textwidth}{Mental health }  &    -0.019 &    -0.085 &     0.027 &     0.021 &    -0.023 &    -0.039 &    -0.059 \\ 
 & (    0.080)  & (    0.132)  & (    0.269)  & (    0.106)  & (    0.091)  & (    0.084)  & (    0.109)  \\ \addlinespace[1ex] 
\multirow[t]{ 2 }{0.2\textwidth}{Beneficiary-specific consumption }  &    -0.179 &    -0.055 &    -0.305 &    -0.229 &    -0.219 &    -0.152 &    -0.099 \\ 
 & (    0.125)  & (    0.192)  & (    0.417)  & (    0.181)  & (    0.138)  & (    0.128)  & (    0.160)  \\
\bottomrule
\end{tabular}
\floatfoot{
	Notes:  Table reports the coefficient on the differential effect of HD over cost-equivalent cash using seven different specifications.  Column 1 is the linear adjustment reported elsewhere.  Column 2 includes a quadratic, and column 3 a quadratic and cubic term in the cost deviations from Gikuriro.  Columns 4-7 leave out one of the cash treatment arms and repeat the linear cost adjustment.  Asterices denote significance at the 10, 5, and 1 percent levels, and are based on household-clustered standard errors, in parentheses. 
}
\end{footnotesize}
\end{table}

\begin{table}[h]\centering
\caption{{\bf Benefit-Cost Ratios, Secondary Outcomes}.}
     \begin{footnotesize}
 \label{t:benefit_cost_ratios_secondary}
 \begin{tabular}{l *{6}{S[table-format=03.3]} SSSS}
\toprule
 & & \multicolumn{4}{c}{GiveDirectly} & & \multicolumn{4}{c}{$ p$-values} \\ 
\cmidrule(lr){3-6} \cmidrule(lr){8-11}
\multicolumn{1}{c}{\text{ }} & \multicolumn{1}{c}{\text{HD}} & \multicolumn{1}{c}{\text{Lower}} & \multicolumn{1}{c}{\text{Middle}} & \multicolumn{1}{c}{\text{Upper}} & \multicolumn{1}{c}{\text{Large}} & \multicolumn{1}{c}{\text{Combined}} & \multicolumn{1}{c}{\text{(a)}} & \multicolumn{1}{c}{\text{(b)}} & \multicolumn{1}{c}{\text{(c)}} & \multicolumn{1}{c}{\text{(d)}}\\
\midrule
\multicolumn{11}{l}{\emph{Panel A. Beneficiary welfare}}  \\ 
\addlinespace[1ex] \multirow[t]{ 2 }{0.25\textwidth}{Subjective well-being }  &     0.036 &     0.043 &     0.058 &     0.067 &     0.019 &     0.033 &      0.01 &      0.80 &      0.29 &      0.21 \\ 
 & (    0.021)  & (    0.023)  & (    0.021)  & (    0.014)  & (    0.010)  & (    0.010)  \\ \addlinespace[1ex] 
\multirow[t]{ 2 }{0.25\textwidth}{Mental health }  &     0.019 &     0.026 &     0.022 &     0.006 &     0.017 &    -0.001 &      0.89 &      0.78 &      0.67 &      0.14 \\ 
 & (    0.020)  & (    0.022)  & (    0.018)  & (    0.015)  & (    0.010)  & (    0.010)  \\ \addlinespace[1ex] 
\multirow[t]{ 2 }{0.25\textwidth}{Beneficiary-specific consumption }  &     0.034 &     0.059 &     0.042 &     0.055 &     0.013 &     0.033 &      0.43 &      0.46 &      0.14 &      0.25 \\
\bottomrule
\end{tabular}
\floatfoot{
      Note:  Table gives the impact per \$100 spent, which is calculated by dividing the estimated ITT impacts by the cost per arm in hundreds of dollars.  The standard errors in the table are similarly the ITT SEs divided by costs.  Reported $p$-values in final three columns derived from $F$-tests of hypotheses that cost-benefit ratios are equal between: (a) joint test across all arms, (b) GD Lower and HD; (c) GD Lower and GD Large; and (d) GD Large and Combined arms.
}
\end{footnotesize}
\end{table}

\end{landscape}

\cleardoublepage

\begin{table}[h]\centering
\caption{{\bf Heterogeneity Analysis by Gender}.}
     \begin{footnotesize}
 \label{t:heterogeneity_female}
 \begin{tabular}{l *{5}{S}}
\toprule
\multicolumn{1}{c}{\text{ }} & \multicolumn{1}{c}{\text{Employed}} & \multicolumn{1}{c}{\text{\makecell[b]{Productive\\Hours}}} & \multicolumn{1}{c}{\text{\makecell[b]{Monthly\\Income}}} & \multicolumn{1}{c}{\text{\makecell[b]{Productive\\Assets}}} & \multicolumn{1}{c}{\text{Consumption}}\\
\midrule
\multirow[t]{ 2}{0.2\textwidth}{HD} &      0.03 &      4.78 &      0.37 &      1.47\ensuremath{^{*}} &      0.07 \\ 
 & (0.05)  & (2.73)  & (0.33)  & (0.61)  & (0.10)  \\ 
 & [    1.00]  & [    0.36]  & [    1.00]  & [    0.07]  & [    1.00]  \\ 
\addlinespace[1ex] \multirow[t]{ 2}{0.2\textwidth}{GD main} &     -0.03 &      2.19 &      0.10 &      2.76\ensuremath{^{***}} &      0.33\ensuremath{^{**}} \\ 
 & (0.05)  & (2.78)  & (0.35)  & (0.61)  & (0.11)  \\ 
 & [    1.00]  & [    1.00]  & [    1.00]  & [    0.00]  & [    0.01]  \\ 
\addlinespace[1ex] \multirow[t]{ 2}{0.2\textwidth}{GD large} &     -0.02 &      3.25 &      0.37 &      2.86\ensuremath{^{***}} &      0.14 \\
\bottomrule
\end{tabular}
\floatfoot{
Notes: Table presents tests for heterogeneity of treatment effects by Gender.  Uninteracted coefficients in the first four rows give the treatment effect of the program on men, and the next four rows test for the differential effect between women and men. Standard errors are (in soft brackets) are clustered at the household level to reflect the design effect, and $p$-values corrected for False Discovery Rates across all the outcomes in the table are presented in hard brackets.  Stars on coefficient estimates are derived from the FDR-corrected $p$-values, *=10\%, **=5\%, and ***=1\% significance.  $p$-value in the last row from an F-test on whether treatments have a jointly differential effect by gender.  
}
\end{footnotesize}
\end{table}

\begin{table}[h]\centering
\caption{{\bf Heterogeneity Analysis by Age}.}
     \begin{footnotesize}
 \label{t:heterogeneity_older}
 \begin{tabular}{l *{5}{S}}
\toprule
\multicolumn{1}{c}{\text{ }} & \multicolumn{1}{c}{\text{Employed}} & \multicolumn{1}{c}{\text{\makecell[b]{Productive\\Hours}}} & \multicolumn{1}{c}{\text{\makecell[b]{Monthly\\Income}}} & \multicolumn{1}{c}{\text{\makecell[b]{Productive\\Assets}}} & \multicolumn{1}{c}{\text{Consumption}}\\
\midrule
\multirow[t]{ 2}{0.2\textwidth}{HD} &      0.04 &      4.42 &      0.65 &      0.68 &     -0.02 \\ 
 & (0.05)  & (2.48)  & (0.42)  & (0.55)  & (0.10)  \\ 
 & [    1.00]  & [    0.71]  & [    0.75]  & [    1.00]  & [    1.00]  \\ 
\addlinespace[1ex] \multirow[t]{ 2}{0.2\textwidth}{GD main} &      0.00 &      1.63 &      0.02 &      2.77\ensuremath{^{***}} &      0.24 \\ 
 & (0.05)  & (2.40)  & (0.43)  & (0.55)  & (0.10)  \\ 
 & [    1.00]  & [    1.00]  & [    1.00]  & [    0.00]  & [    0.16]  \\ 
\addlinespace[1ex] \multirow[t]{ 2}{0.2\textwidth}{GD large} &      0.05 &      1.70 &      1.29 &      2.56\ensuremath{^{**}} &      0.04 \\
\bottomrule
\end{tabular}
\floatfoot{
Notes: Table presents tests for heterogeneity of treatment effects by age.  First four rows give effect of treatment among young, and next four rows test for differential treatment effect for those 23 and over. Standard errors are (in soft brackets) are clustered at the household level to reflect the design effect, and $p$-values corrected for False Discovery Rates across all the outcomes in the table are presented in hard brackets.  Stars on coefficient estimates are derived from the FDR-corrected $p$-values, *=10\%, **=5\%, and ***=1\% significance.  $p$-value in the last row from an F-test on whether treatments have a jointly differential effect by gender. 
}
\end{footnotesize}
\end{table}

\begin{table}[h]\centering
\caption{{\bf Heterogeneity Analysis by Baseline Consumption}.}
     \begin{footnotesize}
 \label{t:heterogeneity_consump}
 \begin{tabular}{l *{5}{S}}
\toprule
\multicolumn{1}{c}{\text{ }} & \multicolumn{1}{c}{\text{Employed}} & \multicolumn{1}{c}{\text{\makecell[b]{Productive\\Hours}}} & \multicolumn{1}{c}{\text{\makecell[b]{Monthly\\Income}}} & \multicolumn{1}{c}{\text{\makecell[b]{Productive\\Assets}}} & \multicolumn{1}{c}{\text{Consumption}}\\
\midrule
\multirow[t]{ 2}{0.2\textwidth}{HD} &      0.03 &      3.66 &      0.16 &      0.84 &      0.07 \\ 
 & (0.03)  & (1.60)  & (0.27)  & (0.37)  & (0.06)  \\ 
 & [    0.90]  & [    0.13]  & [    1.00]  & [    0.13]  & [    0.87]  \\ 
\addlinespace[1ex] \multirow[t]{ 2}{0.2\textwidth}{GD main} &      0.00 &      1.62 &      0.09 &      2.15\ensuremath{^{***}} &      0.20\ensuremath{^{**}} \\ 
 & (0.03)  & (1.59)  & (0.27)  & (0.37)  & (0.07)  \\ 
 & [    1.00]  & [    0.87]  & [    1.00]  & [    0.00]  & [    0.03]  \\ 
\addlinespace[1ex] \multirow[t]{ 2}{0.2\textwidth}{GD large} &      0.02 &      2.28 &      0.65 &      2.83\ensuremath{^{***}} &      0.17 \\
\bottomrule
\end{tabular}
\floatfoot{
Notes: Table presents tests for heterogeneity of treatment effects by baseline Household Consumption.  Consumption demeaned before interaction so first four rows give effect of treatment at average value, and next four rows test for differential treatment effect by consumption. Standard errors are (in soft brackets) are clustered at the household level to reflect the design effect, and $p$-values corrected for False Discovery Rates across all the outcomes in the table are presented in hard brackets.  Stars on coefficient estimates are derived from the FDR-corrected $p$-values, *=10\%, **=5\%, and ***=1\% significance.  $p$-value in the last row from an F-test on whether treatments have a jointly differential effect by gender.  
}
\end{footnotesize}
\end{table}

\begin{table}[h]\centering
\caption{{\bf Heterogeneity Analysis by Sector-level Employment}.}
     \begin{footnotesize}
 \label{t:heterogeneity_employ}
 \begin{tabular}{l *{5}{S}}
\toprule
\multicolumn{1}{c}{\text{ }} & \multicolumn{1}{c}{\text{Employed}} & \multicolumn{1}{c}{\text{\makecell[b]{Productive\\Hours}}} & \multicolumn{1}{c}{\text{\makecell[b]{Monthly\\Income}}} & \multicolumn{1}{c}{\text{\makecell[b]{Productive\\Assets}}} & \multicolumn{1}{c}{\text{Consumption}}\\
\midrule
\multirow[t]{ 2}{0.2\textwidth}{HD} &      0.02 &      3.41 &      0.14 &      0.79 &      0.04 \\ 
 & (0.03)  & (1.61)  & (0.27)  & (0.37)  & (0.06)  \\ 
 & [    1.00]  & [    0.25]  & [    1.00]  & [    0.25]  & [    1.00]  \\ 
\addlinespace[1ex] \multirow[t]{ 2}{0.2\textwidth}{GD main} &      0.00 &      1.54 &      0.08 &      2.10\ensuremath{^{***}} &      0.18 \\ 
 & (0.03)  & (1.60)  & (0.27)  & (0.38)  & (0.07)  \\ 
 & [    1.00]  & [    1.00]  & [    1.00]  & [    0.00]  & [    0.10]  \\ 
\addlinespace[1ex] \multirow[t]{ 2}{0.2\textwidth}{GD large} &      0.02 &      2.15 &      0.63 &      2.78\ensuremath{^{***}} &      0.14 \\
\bottomrule
\end{tabular}
\floatfoot{
Notes: Table presents tests for heterogeneity of treatment effects by baseline Employment Rates. Employment demeaned before interaction so first four rows give effect of treatment at average value, and next four rows test for differential treatment effect by employment rates. Standard errors are (in soft brackets) are clustered at the household level to reflect the design effect, and $p$-values corrected for False Discovery Rates across all the outcomes in the table are presented in hard brackets.  Stars on coefficient estimates are derived from the FDR-corrected $p$-values, *=10\%, **=5\%, and ***=1\% significance.  $p$-value in the last row from an F-test on whether treatments have a jointly differential effect by gender.  
}
\end{footnotesize}
\end{table}

\cleardoublepage

\begin{table}[h]\centering
\caption{{\bf Simple Spillover Analysis}.}
     \begin{footnotesize}
 \label{t:saturation_levels}
 \begin{tabular}{l *{5}{S}}
\toprule
\multicolumn{1}{c}{\text{ }} & \multicolumn{1}{c}{\text{Employed}} & \multicolumn{1}{c}{\text{\makecell[b]{Productive\\Hours}}} & \multicolumn{1}{c}{\text{\makecell[b]{Monthly\\Income}}} & \multicolumn{1}{c}{\text{\makecell[b]{Productive\\Assets}}} & \multicolumn{1}{c}{\text{Consumption}}\\
\midrule
\multirow[t]{ 2}{0.2\textwidth}{HD} &      0.02 &      2.70 &      0.14 &      0.93\ensuremath{^{**}} &      0.03 \\ 
 & (0.03)  & (1.24)  & (0.20)  & (0.31)  & (0.05)  \\ 
 & [    0.69]  & [    0.12]  & [    0.69]  & [    0.02]  & [    0.69]  \\ 
\addlinespace[1ex] \multirow[t]{ 2}{0.2\textwidth}{GD main} &     -0.01 &      0.07 &      0.11 &      2.18\ensuremath{^{***}} &      0.16\ensuremath{^{**}} \\ 
 & (0.03)  & (1.27)  & (0.20)  & (0.33)  & (0.05)  \\ 
 & [    0.69]  & [    0.91]  & [    0.69]  & [    0.00]  & [    0.01]  \\ 
\addlinespace[1ex] \multirow[t]{ 2}{0.2\textwidth}{GD Huge treatment} &      0.02 &      2.08 &      0.69 &      2.97\ensuremath{^{***}} &      0.17 \\
\bottomrule
\end{tabular}
\floatfoot{
Notes:  Table analyzes spillover effects of the three main treatments (HD, GD Main, and GD Large) on the five primary outcomes.  The first three rows are dummy variables for own treatment status, and the next three are the saturation rates for the three treatments among others in the village, so measure the marginal effect of going from no one else treated to everyone else treated.  Regressions include but do not report the lagged dependent variable, fixed effects for randomization blocks, and a set of LASSO-selected baseline covariates, and are weighted to reflect intensive tracking.  Standard errors are (in soft brackets) are clustered at the household level to reflect the design effect, and $p$-values corrected for False Discovery Rates across all the outcomes in the table are presented in hard brackets.  Stars on coefficient estimates are derived from the FDR-corrected $p$-values, *=10\%, **=5\%, and ***=1\% significance.  Bottom row is the $p$-value on an F-test of the joint significance of the three saturation terms.
}
\end{footnotesize}
\end{table}

\begin{table}[h]\centering
\caption{{\bf Spillovers on Household Consumption}.}
     \begin{footnotesize}
 \label{t:interference_hh_month_consumption_pc}
 \begin{tabular}{l *{3}{S}}
\toprule
 & \multicolumn{3}{c}{Treatment} \\ 
\cmidrule(lr){2-4} 
\multicolumn{1}{c}{\text{ }} & \multicolumn{1}{c}{\text{HD}} & \multicolumn{1}{c}{\text{GD Main}} & \multicolumn{1}{c}{\text{GD Huge}}\\
\midrule
\multicolumn{4}{l}{\emph{Direct effects of treatment at saturation level of zero}}  \\ 
\addlinespace[1ex] \multirow[t]{ 2}{0.2\textwidth}{Direct effect} &     -0.14 &     -0.04 &      0.45 \\ 
 & (0.13)  & (0.14)  & (0.26)  \\ 
 & [    0.30]  & [    0.60]  & [    0.16]  \\ 
\addlinespace[1ex] \multicolumn{4}{l}{\emph{Spillover effects of treatment onto control individuals}}  \\ 
\addlinespace[1ex] \multirow[t]{ 2}{0.2\textwidth}{Spillover to control} &     -0.26 &     -0.52\ensuremath{^{*}} &     -0.52 \\ 
 & (0.19)  & (0.18)  & (0.28)  \\ 
 & [    0.23]  & [    0.05]  & [    0.15]  \\ 
\addlinespace[1ex] \multicolumn{4}{l}{\emph{Additional effect of treatment onto individuals assigned to\ldots}}  \\ 
\addlinespace[1ex] \multirow[t]{ 2}{0.2\textwidth}{HD} &      0.15 &      0.40 &     -0.28 \\ 
 & (0.21)  & (0.21)  & (0.37)  \\ 
 & [    0.42]  & [    0.15]  & [    0.42]  \\ 
\addlinespace[1ex] \multirow[t]{ 2}{0.2\textwidth}{GD main} &     -0.16 &      0.49 &      0.78 \\ 
 & (0.23)  & (0.22)  & (0.34)  \\ 
 & [    0.42]  & [    0.14]  & [    0.14]  \\ 
\addlinespace[1ex] \multirow[t]{ 2}{0.2\textwidth}{GD large} &     -0.79 &      0.20 &     -0.80 \\ 
 & (0.41)  & (0.36)  & (0.75)  \\ 
 & [    0.15]  & [    0.49]  & [    0.30]  \\ 
\addlinespace[1ex] Saturation mean  &      0.36 &      0.36 &      0.09 \\ 
Saturation SD  &      0.23 &      0.23 &      0.13 \\ 
$ p$-value  &      0.01 &      0.03 &      0.02 \\ 
\addlinespace[1ex] 
\bottomrule
\end{tabular}
\floatfoot{
Notes: Each column describes the direct and spillover effects of a specific treatment on Household Consumption (IHS); all results in the table are from a single estimation. Saturation mean and standard deviation correspond to the distribution of saturation rates for the treatment in question. Standard errors are (in soft brackets) are clustered at the household level to reflect the design effect, and $p$-values corrected for False Discovery Rates across all the outcomes in the table are presented in hard brackets.  Stars on coefficient estimates are derived from the FDR-corrected $p$-values, *=10\%, **=5\%, and ***=1\% significance.  $p$-value in the last row corresponds to a test for whether the treatment in question has interference effects on any arm, including control.
}
\end{footnotesize}
\end{table}

\begin{table}[h]\centering
\caption{{\bf Spillovers on Employment}.}
     \begin{footnotesize}
 \label{t:interference_bn_employed}
 \begin{tabular}{l *{3}{S}}
\toprule
 & \multicolumn{3}{c}{Treatment} \\ 
\cmidrule(lr){2-4} 
\multicolumn{1}{c}{\text{ }} & \multicolumn{1}{c}{\text{HD}} & \multicolumn{1}{c}{\text{GD Main}} & \multicolumn{1}{c}{\text{GD Huge}}\\
\midrule
\multicolumn{4}{l}{\emph{Direct effects of treatment at saturation level of zero}}  \\ 
\addlinespace[1ex] \multirow[t]{ 2}{0.2\textwidth}{Direct effect} &     -0.05 &     -0.07 &     -0.11 \\ 
 & (0.08)  & (0.07)  & (0.12)  \\ 
 & [    1.00]  & [    0.98]  & [    0.98]  \\ 
\addlinespace[1ex] \multicolumn{4}{l}{\emph{Spillover effects of treatment onto control individuals}}  \\ 
\addlinespace[1ex] \multirow[t]{ 2}{0.2\textwidth}{Spillover to control} &     -0.13 &     -0.16 &      0.14 \\ 
 & (0.09)  & (0.09)  & (0.15)  \\ 
 & [    0.98]  & [    0.98]  & [    0.98]  \\ 
\addlinespace[1ex] \multicolumn{4}{l}{\emph{Additional effect of treatment onto individuals assigned to\ldots}}  \\ 
\addlinespace[1ex] \multirow[t]{ 2}{0.2\textwidth}{HD} &      0.13 &      0.10 &     -0.12 \\ 
 & (0.12)  & (0.11)  & (0.19)  \\ 
 & [    0.98]  & [    0.98]  & [    1.00]  \\ 
\addlinespace[1ex] \multirow[t]{ 2}{0.2\textwidth}{GD main} &      0.03 &      0.19 &     -0.22 \\ 
 & (0.12)  & (0.12)  & (0.19)  \\ 
 & [    1.00]  & [    0.98]  & [    0.98]  \\ 
\addlinespace[1ex] \multirow[t]{ 2}{0.2\textwidth}{GD large} &      0.35 &      0.06 &     -0.13 \\ 
 & (0.18)  & (0.18)  & (0.39)  \\ 
 & [    0.98]  & [    1.00]  & [    1.00]  \\ 
\addlinespace[1ex] Saturation mean  &      0.36 &      0.36 &      0.09 \\ 
Saturation SD  &      0.23 &      0.23 &      0.13 \\ 
$ p$-value  &      0.33 &      0.40 &      0.84 \\ 
\addlinespace[1ex] 
\bottomrule
\end{tabular}
\floatfoot{
Notes: Each column describes the direct and spillover effects of a specific treatment on Employment; all results in the table  are from a single estimation. Saturation mean and standard deviation correspond to the distribution of saturation rates for the treatment in question.  Regressions include but do not report the lagged dependent variable, fixed effects for randomization blocks, and a set of LASSO-selected baseline covariates, and are weighted to reflect intensive tracking.  Standard errors are (in soft brackets) are clustered at the household level to reflect the design effect, and $p$-values corrected for False Discovery Rates across all the outcomes in the table are presented in hard brackets.  Stars on coefficient estimates are derived from the FDR-corrected $p$-values, *=10\%, **=5\%, and ***=1\% significance.  $p$-value in the last row corresponds to a test for whether the treatment in question has interference effects on any arm, including control.  
}
\end{footnotesize}
\end{table}

\begin{table}[h]\centering
\caption{{\bf Spillovers on Monthly Income}.}
     \begin{footnotesize}
 \label{t:interference_bn_monthly_income}
 \begin{tabular}{l *{3}{S}}
\toprule
 & \multicolumn{3}{c}{Treatment} \\ 
\cmidrule(lr){2-4} 
\multicolumn{1}{c}{\text{ }} & \multicolumn{1}{c}{\text{HD}} & \multicolumn{1}{c}{\text{GD Main}} & \multicolumn{1}{c}{\text{GD Huge}}\\
\midrule
\multicolumn{4}{l}{\emph{Direct effects of treatment at saturation level of zero}}  \\ 
\addlinespace[1ex] \multirow[t]{ 2}{0.2\textwidth}{Direct effect} &     -0.76 &     -0.00 &     -0.69 \\ 
 & (0.57)  & (0.57)  & (0.96)  \\ 
 & [    0.74]  & [    1.00]  & [    0.74]  \\ 
\addlinespace[1ex] \multicolumn{4}{l}{\emph{Spillover effects of treatment onto control individuals}}  \\ 
\addlinespace[1ex] \multirow[t]{ 2}{0.2\textwidth}{Spillover to control} &     -1.00 &     -0.90 &     -1.16 \\ 
 & (0.72)  & (0.70)  & (1.21)  \\ 
 & [    0.74]  & [    0.74]  & [    0.74]  \\ 
\addlinespace[1ex] \multicolumn{4}{l}{\emph{Additional effect of treatment onto individuals assigned to\ldots}}  \\ 
\addlinespace[1ex] \multirow[t]{ 2}{0.2\textwidth}{HD} &      1.39 &      0.47 &      2.36 \\ 
 & (0.91)  & (0.85)  & (1.48)  \\ 
 & [    0.74]  & [    0.74]  & [    0.74]  \\ 
\addlinespace[1ex] \multirow[t]{ 2}{0.2\textwidth}{GD main} &     -0.36 &      0.48 &      0.70 \\ 
 & (0.90)  & (0.90)  & (1.54)  \\ 
 & [    0.74]  & [    0.74]  & [    0.74]  \\ 
\addlinespace[1ex] \multirow[t]{ 2}{0.2\textwidth}{GD large} &      1.99 &      0.54 &      5.08 \\ 
 & (1.54)  & (1.24)  & (2.74)  \\ 
 & [    0.74]  & [    0.74]  & [    0.74]  \\ 
\addlinespace[1ex] Saturation mean  &      0.36 &      0.36 &      0.09 \\ 
Saturation SD  &      0.23 &      0.23 &      0.13 \\ 
$ p$-value  &      0.35 &      0.74 &      0.24 \\ 
\addlinespace[1ex] 
\bottomrule
\end{tabular}
\floatfoot{
Notes: Each column describes the direct and spillover effects of a specific treatment on Monthly Income (IHS); all results in the table  are from a single estimation. Saturation mean and standard deviation correspond to the distribution of saturation rates for the treatment in question.  Regressions include but do not report the lagged dependent variable, fixed effects for randomization blocks, and a set of LASSO-selected baseline covariates, and are weighted to reflect intensive tracking. Standard errors are (in soft brackets) are clustered at the household level to reflect the design effect, and $p$-values corrected for False Discovery Rates across all the outcomes in the table are presented in hard brackets.  Stars on coefficient estimates are derived from the FDR-corrected $p$-values, *=10\%, **=5\%, and ***=1\% significance.  $p$-value in the last row corresponds to a test for whether the treatment in question has interference effects on any arm, including control.
}
\end{footnotesize}
\end{table}

\begin{table}[h]\centering
\caption{{\bf Spillovers on Productive Hours}.}
     \begin{footnotesize}
 \label{t:interference_bn_productive_hrs}
 \input{tables/interference_bn_productive_hrs}
\floatfoot{
Notes: Each column describes the direct and spillover effects of a specific treatment on Productive Hours; all results in the table  are from a single estimation. Saturation mean and standard deviation correspond to the distribution of saturation rates for the treatment in question.  Regressions include but do not report the lagged dependent variable, fixed effects for randomization blocks, and a set of LASSO-selected baseline covariates, and are weighted to reflect intensive tracking. Standard errors are (in soft brackets) are clustered at the household level to reflect the design effect, and $p$-values corrected for False Discovery Rates across all the outcomes in the table are presented in hard brackets.  Stars on coefficient estimates are derived from the FDR-corrected $p$-values, *=10\%, **=5\%, and ***=1\% significance.   $p$-value in the last corresponds to a test for whether the treatment in question has interference effects on any arm, including control.
}
\end{footnotesize}
\end{table}

\begin{table}[h]\centering
\caption{{\bf Spillovers on Productive Asset Values}.}
     \begin{footnotesize}
 \label{t:interference_bn_tot_prod_assetval}
 \input{tables/interference_bn_tot_prod_assetval}
\floatfoot{
Notes: Each column describes the direct and spillover effects of a specific treatment on Productive Assets (IHS); all results in the table  are from a single estimation. Saturation mean and standard deviation correspond to the distribution of saturation rates for the treatment in question.  Regressions include but do not report the lagged dependent variable, fixed effects for randomization blocks, and a set of LASSO-selected baseline covariates, and are weighted to reflect intensive tracking. Standard errors are (in soft brackets) are clustered at the household level to reflect the design effect, and $p$-values corrected for False Discovery Rates across all the outcomes in the table are presented in hard brackets.  Stars on coefficient estimates are derived from the FDR-corrected $p$-values, *=10\%, **=5\%, and ***=1\% significance.  $p$-value in the last row corresponds to a test for whether the treatment in question has interference effects on any arm, including control.
}
\end{footnotesize}
\end{table}

\cleardoublepage

\begin{table}
\caption{{\bf Are treatments protective against measured Covid shocks?}}
\label{t:tx_protect}
\centering
\begin{tabular}{l *{5}{c}}
\toprule
\multicolumn{1}{c}{\text{ }} & \multicolumn{1}{c}{\text{Employed}} & \multicolumn{1}{c}{\text{Productive hours}} & \multicolumn{1}{c}{\text{Income}} & \multicolumn{1}{c}{\text{Productive assets}} & \multicolumn{1}{c}{\text{Consumption}}\\
\midrule
HD  &      0.02 &      3.52\ensuremath{^{**}} &      0.16 &      0.82\ensuremath{^{**}} &      0.05 \\ 
 & (0.03)  & (1.63)  & (0.26)  & (0.37)  & (0.06)  \\ 
\addlinespace[1ex] \quad$\times$ index  &      0.15 &     -0.17 &      0.29 &      0.12 &      0.46 \\ 
 & (0.32)  & (0.31)  & (0.24)  & (0.26)  & (0.28)  \\ 
\addlinespace[1ex] GD main  &      0.00 &      1.48 &      0.10 &      2.14\ensuremath{^{***}} &      0.17\ensuremath{^{***}} \\ 
 & (0.03)  & (1.64)  & (0.26)  & (0.37)  & (0.06)  \\ 
\addlinespace[1ex] \quad$\times$ index  &      0.17 &      0.38 &     -0.12 &      0.33 &      0.19 \\
\bottomrule
\end{tabular}
\floatfoot{
	\begin{footnotesize}
	Notes: This table proceeds in the following steps.  First, we use second-order polynomials in the cumulative shock indices to predict endline outcomes in the control group;  following \citet{AbaChiWes18restat}, we use a leave-one-out approach to omit each control-group observation from the regression on which its prediction is based. We then use predicted endline outcomes in the control group to predict counterfactual endline outcomes in other treatment arms, and interact the centered predictions with treatment indicators.  All specifications include controls for the baseline value of the outcome, as well as lasso-selected controls and block fixed effects, as in the ITT specification.  \emph{Predictive index} is the predicted value of the endline outcome, based on plausibly exogenous covid shock measures, with model estimated by cross-validated lasso in the control group only.  
	\end{footnotesize}
}

\end{table}

\begin{table}[!hbtp]
\caption{{\bf Association between Covid-19 shock measures and endline outcomes in control group}}
\label{t:control_covid_outcomes}
\centering
\input{tables/control_covid_outcomes}
\floatfoot{ 
\begin{footnotesize}
Note:  Each column within each panel represents a separate regression.  All regressions control for baseline values of the corresponding outcome and for block fixed effects.
\end{footnotesize}
}
\end{table}

\begin{landscape}

\begin{table}[h]\centering
\caption{{\bf ITT Effects on Transfers}.}
     \begin{footnotesize}
 \label{t:itt_transfers}
 \input{tables/itt_transfers}
\floatfoot{
      Note:  Regressions include but do not report the lagged dependent variable, fixed effects for randomization blocks, and a set of LASSO-selected baseline covariates, and are weighted to reflect intensive tracking. Standard errors are (in soft brackets) are clustered at the household level to reflect the design effect, and $p$-values corrected for False Discovery Rates across all the outcomes in the table are presented in hard brackets.  Stars on coefficient estimates are derived from the FDR-corrected $p$-values, *=10\%, **=5\%, and ***=1\% significance. Reported $p$-values in final three columns derived from $F$-tests of hypotheses that cost-benefit ratios are equal between: (a) GD Lower and HD; (b) GD Lower and GD Large; and (c) GD Large and Combined treatments. 
}
\end{footnotesize}
\end{table}

\cleardoublepage

\end{landscape}

\end{document}